\documentclass[3p,twocolumn]{elsarticle}

\usepackage[utf8]{inputenc}
\usepackage{multirow}
\usepackage{multicol}
\usepackage{subcaption} 
\usepackage{graphicx} 
\usepackage{circuitikz} 
\tikzset{>=latex} 
\usepackage{amsmath} 
\usepackage{amssymb}
\usepackage{cuted}
\usepackage{url}
\usepackage{lineno}
\usepackage{changepage}
\usepackage[linesnumbered,ruled]{algorithm2e}
\usepackage{graphicx}
\usepackage{bm}
\usepackage{amsmath}      
\usepackage{algpseudocode}  
\usepackage{float}        
\usepackage{amsfonts}     
\usepackage[utf8]{inputenc}
\usepackage{booktabs}   
\usepackage{tabularx}   
\usepackage{booktabs}       
\usepackage{tabularx}       
\usepackage{caption}        
\usepackage{makecell}       
\usepackage{threeparttable} 
\usepackage{rotating}       
\usepackage[utf8]{inputenc}
\usepackage{booktabs}   
\usepackage{caption}    
\usepackage{graphicx}   
\usepackage{array}      
\usepackage{comment}

\newcolumntype{P}[1]{>{\raggedright\arraybackslash}p{#1}}

\newcommand{\simm}{\raise.17ex\hbox{$\scriptstyle\sim$}}
\newcommand{\bmat}{\begin{bmatrix}}
\newcommand{\emat}{\end{bmatrix}}

\usepackage{amsmath}
\usepackage{amssymb}
\usepackage{array} 
\usepackage{booktabs} 
\newcommand{\new}[1]{\textcolor{black}{#1}}

\journal{Applied Energy}

\begin{document}
\begin{frontmatter}

\title{Comparative Field Deployment of Model-Based Reinforcement Learning and Model Predictive Control for Residential HVAC}

\author[1,2]{Ozan Baris Mulayim}
\author[3]{Elias N. Pergantis}
\author[4]{Levi D. Reyes Premer}
\author[5]{Bingqing Chen}
\author[1]{Guannan Qu}
\author[4]{Kevin J. Kircher}
\author[1,2]{Mario Bergés\corref{correspondent}\fnref{amazon}}

\address[1]{College of Engineering, Carnegie Mellon University, 5000 Forbes Ave, Pittsburgh, PA 15213, USA}
\address[2]{Wilton E. Scott Institute for Energy Innovation, Carnegie Mellon University, 5000 Forbes Ave, Pittsburgh, PA 15213, USA}
\address[3]{Trane Technologies, Residential R\&D Group, 6200 Troup Hwy, Tyler, TX 75707, USA}
\address[4]{College of Engineering, Purdue University, 610 Purdue Mall, West Lafayette, IN 47907, USA}
\address[5]{Bosch Center for Artificial Intelligence; Pittsburgh, PA, USA}
\cortext[correspondent]{Corresponding author: \texttt{marioberges@cmu.edu}}

\begin{abstract}

Model Predictive Control (MPC) has demonstrated significant performance improvements over today's control methods for residential Heating, Ventilation, and Air Conditioning (HVAC), but deploying MPC often requires substantial engineering effort. Reinforcement Learning (RL) may offer comparable performance with easier deployment, but its practical application for residential HVAC remains largely undemonstrated, leaving open questions related to occupant comfort and data requirements. To investigate these issues, we deployed one MPC variant and one model-based RL variant for one month each in an occupied house in a cold climate. The controllers adjusted an air-to-air heat pump's thermostat temperature setpoint based on measurements of the indoor temperature and the electric power used for heating. Relative to constant-setpoint operation, MPC saved 18.1\% (95\% confidence interval: 4.4 to 30.9\%) of weather-normalized heat pump energy and RL saved 20.9\% (2.6 to 38.3\%). MPC maintained acceptable occupant comfort. RL kept the house cooler, particularly during an initial adaptation phase, leading to three reports of occupant discomfort. The two algorithms had similar data requirements. We estimate that for a fresh deployment in another house, RL would take about one-third less engineering effort than MPC. While RL reduces deployment effort, it faces difficulties related to safe controller initialization and to mismatches between the modeled and true state and action spaces.

\end{abstract}


\end{frontmatter}

\section{Introduction}

Advanced control strategies for Heating, Ventilation and Air Conditioning (HVAC) systems have demonstrated significant potential for improving energy efficiency and occupant comfort \cite{li2014review, bengea2012model}. Two prominent approaches are Model Predictive Control (MPC) and Reinforcement Learning (RL), each offering distinct advantages and facing unique challenges. MPC uses a mathematical model of the building's thermal dynamics to predict future states and optimize control actions over a receding horizon. The model may be physics-based, data-driven, or hybrid, and may or may not adapt online to time-varying dynamics. MPC's core strengths lie in its  predictive capability, allowing for proactive load-shifting or pre-cooling/heating, and its ability to handle constraints such as temperature bounds or equipment limits. Numerous MPC implementations demonstrate its effectiveness in optimizing building operations (e.g., \cite{khabbazi2025lessons}).

RL offers an alternative paradigm where control policies are learned through trial-and-error interactions with the environment, guided by a reward signal designed to encapsulate control objectives \cite{sutton1998reinforcement}. The primary appeal of RL lies in its potential to automatically discover control strategies without requiring an explicit system model, and its inherent capacity for continuous adaptation to changing conditions through ongoing learning \cite{chen_gnu-rl_2019}. Relative to MPC, RL could reduce modeling effort and improve robustness to system variations over time. However, translating RL from simulation to reliable real-world implementation faces practical hurdles. Key challenges include ensuring operational safety and occupant comfort during the exploration phase necessary for learning, the often substantial amount of interaction data required to converge to effective policies (sample efficiency), and the potential difficulty in interpreting learned policies or rigorously guaranteeing constraint satisfaction \cite{dulac2021challenges}. 

An often overlooked barrier to RL deployment is the reliance on training environments. Most proposed RL variants depend on pre-training within high-fidelity simulators \cite{chen_gnu-rl_2019}. However, developing a high-fidelity simulator for every specific building is a labor-intensive task that effectively reintroduces the engineering bottleneck that RL aims to eliminate. Therefore, validating ``simulator-free'' approaches --- where agents learn directly from historical or online data --- is essential for proving RL's viability as a scalable solution. Consequently, while promising, RL research in HVAC control remains largely confined to simulation studies, with only a handful of real-world deployments documented in residential settings \citep{leurs2016beyond, kurte2020evaluating, svetozarevic2022data, montazeri2025fully}. The combined duration of all peer-reviewed field experiments in residential buildings of which the authors are aware totals merely 43 days, highlighting a substantial gap between theoretical promise and demonstrated, practical application.

Given the distinct strengths and challenges inherent to both MPC and RL, understanding their practical trade-offs in real-world settings is crucial. There is a need for comparative studies that evaluate not just performance metrics but also the deployment effort, adaptability, and operational robustness  over extended periods in real buildings.

\begin{figure*}[ht]
    \centering
    \includegraphics[width=1.8\columnwidth]{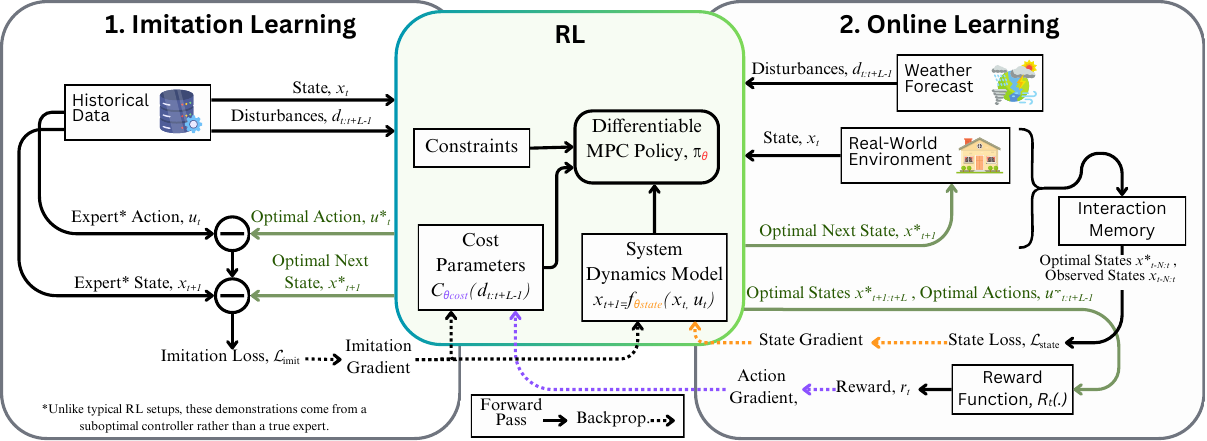} \\
    \vspace{2mm}
    \includegraphics[width=1.8\columnwidth]{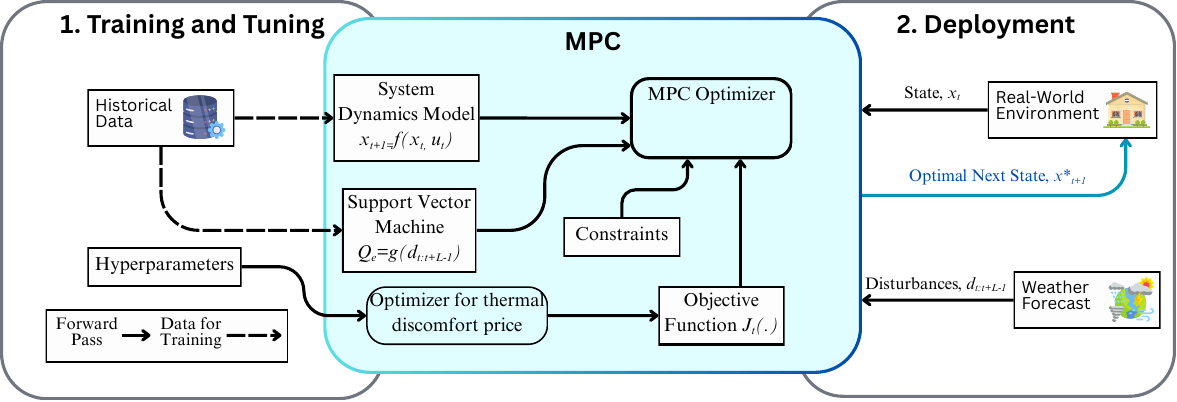} \\
    \caption{Overview of the RL and MPC controllers}
    \label{fig:rlandmpc}
\end{figure*}

Towards addressing this gap, this paper moves beyond simulation \cite{wang2023comparison} to confront the practical challenges and trade-offs of deploying advanced controllers in the real world. We aim to answer questions regarding the balance between the engineering effort of MPC and the safety hurdles of RL \cite{nagy2023ten, wang2023comparison}. To do this, we deploy one MPC variant and one model-based RL variant (illustrated in Figure \ref{fig:rlandmpc}) for one month each in the same occupied house (Figure \ref{fig:dc_house}) in West Lafayette, Indiana, for supervisory control of an air-to-air heat pump. We deploy RL without the safety net of a high-fidelity simulator. The challenge of this approach became immediately apparent; our initial plan to deploy the Gnu-RL framework \cite{chen_gnu-rl_2019} was unworkable due to practical limitations, necessitating significant modifications to ensure stable and effective operation. Thus, we deployed a modified version of Gnu-RL called Ibex-RL \cite{mulayim2025physics}. Ibex-RL learns (1) a physics-informed system dynamics model, and (2) a reward function with minimal user guidance. The MPC variant learns a similar physics-informed system dynamics model, but requires more engineer effort to oversee training and tune cost-function parameters. By comparing MPC and RL in the same real, occupied house, this study offers empirical insights into the trade-offs between deployment effort, adaptability, and performance under similar boundary conditions.

In the field experiments, the MPC and RL variants gave similar weather-normalized heat pump energy savings (18.1\% for MPC, 20.9\% for RL) relative to a constant-setpoint baseline. MPC maintained acceptable occupant comfort, while RL received one discomfort report during an initial adaptation phase shortly after deployment and two discomfort reports midway through the RL testing period. Based on our experience implementing both algorithms, we estimate that a fresh deployment in another house would take about four days for instrumentation and commissioning for either algorithm, plus about five days of control work for MPC and about two days for RL. These estimates are specific to the MPC and RL variants that we deployed. While these variants are arguably well-suited to the residential HVAC control task, other MPC or RL variants might perform differently for energy, comfort, or deployment effort.

In summary, this paper makes three main contributions to the research literature:
\begin{enumerate}
\item The first empirical comparison of RL and MPC for HVAC in an occupied house, providing a residential counterpart to prior commercial building studies \cite{wang2024long}.
\item An analysis of long-term RL performance and adaptation, drawing insights from a month-long residential field test. 
\item A practical roadmap of lessons learned outlining challenges and solutions for model accuracy, safety, and engineering effort when deploying MPC or RL in the wild.
\end{enumerate}
Our goal with this study is to help researchers understand the practical challenges involved in deploying MPC or RL for residential HVAC, highlight research directions that could address these challenges, and ultimately bring the technologies closer to real-world deployability at scale.

The remainder of this paper is organized as follows. Section \ref{sec:related_work} discusses relevant prior work including experimental studies. Section \ref{sec:background} reviews foundational concepts in RL and MPC. Section \ref{sec:approach} details the specific RL and MPC variants that we compare. Section \ref{sec:experiments} details the experimental design. Section \ref{sec:results} presents the empirical results from the field deployments, comparing energy use and occupant comfort under MPC, RL, and a constant-setpoint baseline. Section \ref{sec:discussions} discusses lessons learned and practical improvements for future RL and MPC implementations. Section \ref{sec:conclusions} concludes the paper with a synthesis and a discussion of the limitations and future work.

\section{Related Work}
\label{sec:related_work}
This section reviews prior research on MPC and RL for building HVAC systems. We discuss the state of the art and highlight key methodologies and experimental findings. We identify gaps in the literature, particularly concerning long-term residential deployments and direct comparisons between MPC and RL, motivating the contributions of this paper.

\subsection{Reinforcement Learning}
\label{sec:RLreview}

RL has emerged as a compelling approach for optimizing HVAC systems in buildings, primarily due to its ability to adapt control strategies in response to complex and changing environmental conditions \cite{lymperopoulos2020building}. However, practical deployment necessitates addressing challenges related to learning safety, data efficiency, and interpretability \cite{dulac2021challenges}.

A major hurdle for deploying RL in real buildings is the risk associated with online learning, where an agent explores actions directly within the live environment \cite{dulac2021challenges}. Such exploration can lead to occupant discomfort, energy waste, or equipment strain. This has spurred interest in offline RL, where policies are learned from pre-existing datasets \cite{levine_offline_2020}. While safer, purely offline methods can suffer from limitations in the training data. An \textit{offline pre-training followed by online fine-tuning} strategy balances initial safety with ongoing adaptation \cite{liu2024adaptive}. This hybrid approach has seen success in robotics \cite{nakamoto2024cal, kumar2020conservative} and initial HVAC studies \cite{liu2024adaptive, chen_gnu-rl_2019}.

RL methods can be categorized as model-free or model-based. Model-free RL directly learns control policies or value functions. While deployable (e.g., \cite{zhang2018deep} used a simulation-trained deep RL agent for radiant heating), model-free approaches often struggle with sample efficiency, require high-fidelity simulators, and suffer from distributional shift, potentially limiting their scalability  \cite{li2019transforming, jia2019advanced, levine_offline_2020}. To address training efficiency, a recent work focused on accelerating online learning by integrating expert guidance from abstract models, historical data via offline RL, and predefined rules \cite{xu2025efficient}. However, its approach was only validated in simulated environments, not on a physical building.

Model-based RL aims to improve sample efficiency and planning by first learning a dynamics model from data or simulation. This learned model is then used for control, often via planning methods like MPC. Examples include using neural network dynamics models with MPC variants \cite{zhang2022safe, ding2020mb2c} or Gaussian Process models \cite{an_clue_2023}. However, purely data-driven black-box models, such as deep neural networks, lack physical interpretability and typically require large amount of data, posing challenges for trust and generalizability \cite{nagy2023ten}. 

The boundary between MPC and model-based RL is blurry. Both methods use a model of the system dynamics. MPC traditionally relies on physics-based models that require substantial engineering effort to form and tune. However, data-driven MPC variants learn models through automated system identification procedures, as do most model-based RL variants. Non-adaptive MPC variants, such as the one implemented here \cite{pergantis_field_2024}, use a fixed model and a fixed optimization formulation that are engineered offline. However, adaptive MPC methods update the model or optimization formulation online, as do most model-based RL variants.

While the model-free and model-based RL address specific challenges, integrating adaptability, model-based planning, constraint handling, and end-to-end learning remains crucial for practical HVAC control. In this context, the Differentiable MPC policy \cite{amos_differentiable_2019} offers a compelling direction. This framework allows system dynamics and cost function parameters within an MPC structure to be learned end-to-end. Gnu-RL \cite{chen_gnu-rl_2019} tailored Differentiable MPC to HVAC applications, combining offline imitation learning  with RL-based online adaptation. Gnu-RL supports model-based planning, handles constraints effectively, and allows continuous adaptation, making it arguably well-suited to building energy management.

Despite the promise of RL variants such as Gnu-RL, experimental validation in real buildings, especially residences, is lagging. A recent review of field studies \cite{khabbazi2025lessons} showed that while numerous RL variants have been tested in commercial buildings (e.g., \cite{liu2006experimental, zhang2019whole, chen_gnu-rl_2019, naug2022reinforcement, luo2022controlling, wang2024long, silvestri2025practical}), documented residential RL deployments are scarce, totaling only 43 days of testing across four studies \cite{leurs2016beyond, kurte2020evaluating, svetozarevic2022data, montazeri2025fully}. 

Table \ref{tab:deployments} compares the four prior residential RL deployments and the deployment in this paper. There is a lack of long-term residential tests (the only other month-long evaluation is from \cite{montazeri2025fully}), especially for complex systems like heat pumps, and experimental comparisons between RL and MPC. To the authors' knowledge, this paper is also the first to test a model-based RL variant in an occupied residence. Prior studies used model-free methods, which either require an offline building simulator for training \cite{kurte2020evaluating, svetozarevic2022data, montazeri2025fully} or learn exclusively through online interaction with the real building \cite{leurs2016beyond}.

\begin{table*}[t]
    \centering
    \caption{Overview of Residential Field Deployments (Extended from \cite{khabbazi2025lessons})}
    \label{tab:deployments}
    \resizebox{\textwidth}{!}{%
        \begin{tabular}{@{}P{2.1cm} P{2.1cm} P{2.5cm} P{2.8cm} p{1cm} P{2.2cm} P{2.2cm} P{3.5cm} P{2.5cm} P{3.5cm}@{}}
        \toprule
        \textbf{Study (Year)} & \textbf{Location} & \textbf{Setup} & \textbf{System} & \textbf{Test Days} & \textbf{RL Algorithm} & \textbf{Control Action} &  \textbf{Offline Training Strategy} & \textbf{Objective} & \textbf{Key Findings }\\
        \midrule

        \cite{leurs2016beyond} 2016 & 
        Leuven, Belgium & 
        living lab setup featuring a test room & 
        forced-air integrated with PV system & 
        3 &
        Fitted Q-iteration &
        binary on/off commands & 
        \(\boldsymbol{\times}\) \textbf{No Training} (Learns online via 15 days of interaction with the environment) &
        maximize solar self-utilization & 
        reduced PV peak power injection and synchronized cooling with PV generation vs. measured baseline \\
        \midrule

        \cite{kurte2020evaluating} 2020 & 
        Knoxville, TN, USA & 
        detached energy-efficient house & 
        two-zone air-to-air conditioning (AC) with two-stage compressor and variable-speed fan & 
        5 &
        Deep Q-Network &
        high/low setpoints & 
        \(\boldsymbol{\times}\) \textbf{Simulator-based} (Trained offline with an RC Network simulator) &
        minimize cost & 
        11–21\% cost savings vs. simulated baseline \\
        \midrule

        \cite{svetozarevic2022data} 2022 & 
        D\"ubendorf, Switzerland & 
        residential module in a sustainable demonstrator building & 
        radiant heating with HP and emulated electric vehicle integration & 
        5 &
        Deep Deterministic Policy Gradient &
        continuous radiant floor heating valve modulation & 
        \(\boldsymbol{\times}\) \textbf{Simulator-based} (Trained offline with a Recurrent Neural Network simulator) &
        minimize energy & 
        27\% energy savings vs. measured baseline \\
        \midrule
        
        \cite{montazeri2025fully} 2025 & 
        D\"ubendorf, Switzerland &
        residential module in a sustainable demonstrator building &
        ceiling-embedded radiant heating panels &
        30 &
        Deep Deterministic Policy Gradient &
        temperature change translated into heating valve openings (\%) &
        \(\boldsymbol{\times}\) \textbf{Simulator-based} (Trained offline with a Physically Consistent Neural Network simulator) &
        balance thermal comfort and energy savings &
        26-32\% energy savings without compromising comfort vs. measured baseline\\
        \midrule

        This study 2026 & 
        Lafayette, IN, USA &
        occupied detached house &
        air-to-air heat pump with staged electric resistance backup &
        30 &
        Ibex-RL (model-based) &
        electrical HVAC power translated into thermostat setpoint &
        \(\boldsymbol{\checkmark}\) \textbf{No Simulator} (Learns a policy offline from historical data via Imitation Learning) &
        minimize temperature deviation, total and peak energy use &
        20.9\% energy savings with one-third less deployment effort than MPC\\
        
        \bottomrule
        \end{tabular}%
    }
\end{table*}

\subsection{Model Predictive Control}
\label{sub:mpc_related_work} 

MPC is a mature and widely studied strategy for HVAC control \cite{drgovna2020all}. MPC traditionally uses a model of the system dynamics to predict future states and optimizes a sequence of control actions over a finite horizon to minimize a predefined cost function, subject to operational constraints \cite{kouvaritakis2016model}. The optimization is repeated at each time step based on updated measurements and forecasts. A range of MPC variants extend the basic MPC framework, such as explicitly representing model uncertainty, providing robust or probabilistic guarantees on constraint satisfaction, and continuously adapting system models to time-varying dynamics.

Extensive research exists on MPC for building control, with numerous field demonstrations documented, although challenges remain related to model development and deployment effort, as surveyed in \cite{khabbazi2025lessons}. Examples of MPC field studies span various building types, HVAC systems, and objectives. Residential applications have included controlling radiant floor heating \cite{dong2014real} and hybrid systems \cite{afram2017supervisory}, often focusing on energy or cost minimization under time-varying electricity prices \cite{lindelof2015field}. Other studies have focused on constraint satisfaction such as under whole-home controls \cite{PERGANTIS2025125528} or demand response \cite{KIM2015279}. Commercial building studies have demonstrated MPC on systems like variable air volume \cite{bengea2014implementation}, thermally activated building systems \cite{sturzenegger2015model}, and central chiller plants with thermal storage \cite{ma2011model}, sometimes exploring objectives like demand-side flexibility for grid services \cite{maasoumy2014model}.

This paper builds upon and compares against the MPC implementation detailed in \cite{pergantis_field_2024}. This study addressed several gaps in the literature surveyed in \cite{khabbazi2025lessons}. It provided one of the few long-duration (over one month) MPC field tests in an occupied residence. It focused on a complex but common North American equipment configuration (air-to-air heat pump with staged electric resistance backup) often neglected in prior research. Furthermore, \cite{pergantis_field_2024} incorporated adaptive comfort-cost balancing and demonstrated significant reduction in peak electricity demand.

\subsection{Identifying Research Gaps}

Despite progress in both RL and MPC individually, and techniques aiming to combine them, gaps remain in the experimental literature. As highlighted previously, long-term residential RL field studies are rare compared to commercial deployments \cite{khabbazi2025lessons}. Field comparisons between MPC and RL in residential buildings are also lacking. Simulation studies comparing MPC and RL (e.g., \cite{zhan2023comparing, wang2023comparison, arroyo2022comparison, stoffel2023evaluation}) show conflicting results. One recent work \cite{wang2024long} compared different advanced controllers (i.e., soft actor-critic vs. hierarchical data-driven predictive control vs. differentiable predictive control) in a commercial testbed. Their findings were: (1) a hierarchical data-driven predictive control (their MPC variant) achieved the highest energy savings (over 50\%), followed by RL (48\%), although its performance was sensitive to the specific model structure; (2) controller failures were frequently linked to real-world operational issues like Application Programming Interface (API) communication errors, rather than the core algorithms themselves; and (3) a trade-off existed between online computational cost (highest for their MPC) and offline training time (highest for their RL). \new{Yet, it is not clear if these findings hold for residential deployments.}

\new{This paper aims to address these research gaps. As summarized in the introduction, we claim three main contributions to the research literature on residential HVAC control: The first empirical comparison of MPC and RL, the first long-term field deployment of model-based RL, and a discussion of lessons learned and implementation practicalities.}

\section{Mathematical Background}
\label{sec:background}

This section provides technical background on RL, MPC, and the Differentiable MPC policy framework used in Ibex-RL. While these methods are adaptable to partially-observed systems, this work assumes the state is perfectly observable. Throughout this paper, we adopt standard control theory notation: $x_t \in \mathbb{R}^{n_x}$ denotes the state, $u_t \in \mathbb{R}^{n_u}$ the control action, and $d_t \in \mathbb{R}^{n_d}$ the measurable disturbance at time $t$.

\subsection{Reinforcement Learning Fundamentals}
\label{sub:rl_fundamentals}

RL agents learn control policies by interacting with an environment, modeled as a Markov Decision Process (MDP) \cite{sutton1998reinforcement}. The agent observes $x_t$ and selects $u_t$, receiving a reward $r_{t+1}$ as feedback. The objective is to learn a stationary policy $u_t=\pi(x_t)$ that maximizes the expected value of the return $G_t$, defined as the cumulative discounted reward:
\begin{equation}
G_t = \sum_{k=0}^{\infty} \gamma^k r_{t+k+1} .
\label{eq:generic_return}
\end{equation}
Here $\gamma \in [0, 1)$ is the discount factor balancing immediate and future rewards. 
Many algorithms estimate action-value functions, (\(Q(x_t, u_t) = \mathbb{E}[G_t | x_t, u_t]\)), representing the expected return from taking action $u_t$ in state $x_t$. These are updated iteratively, for example via Q-learning:
\begin{multline}
Q(x_t, u_t) \leftarrow Q(x_t, u_t) \\
+ \alpha [r_{t+1} + \gamma \max_{u'} Q(x_{t+1}, u') - Q(x_t, u_t)] .
\label{eq:q_learning_update}
\end{multline}
Here $\alpha$ is the learning rate and the maximization is performed over all possible actions $u'$ in the next state $x_{t+1}$.

As mentioned earlier, directly applying RL online in buildings is often impractical due to safety and comfort risks during initial exploration. Offline RL, learning from a fixed dataset $\mathcal{D}$, avoids these risks but may yield suboptimal policies \cite{levine_offline_2020}. The hybrid offline-to-online strategy uses offline data for safe initial learning and then uses online interaction for refinement and adaptation, offering a practical compromise \cite{liu2024adaptive}. Gnu-RL \cite{chen_gnu-rl_2019} and Ibex-RL \cite{mulayim2025physics} use the offline-to-online strategy. 

\subsection{Model Predictive Control Fundamentals}
\label{sub:mpc_fundamentals}

MPC uses a model of the system to optimize control actions over a prediction horizon. At each time step $t$, MPC solves a constrained optimization problem to plan a sequence of future control inputs $U^*_t = \{u^*_{t}, \dots, u^*_{t+L-1}\}$ over a receding prediction horizon $L$. This process implicitly predicts the future state trajectory. The system dynamics model is a constraint, linking the control actions to their resulting states. Based on the current state $x_t$ and predicted disturbances $d_{t}, \dots, d_{t+L-1}$, MPC finds a control sequence that minimizes a cost function $J(U_t,x_t)$. The cost function quantifies objectives like energy cost and comfort. An example cost function is 
    \begin{multline}
        \sum_{\ell=0}^{L-1} \left( ||x_{t+\ell+1} - x_{target,t+\ell+1}||^2 + ||u_{t+\ell}||^2 \right) .
        \label{eq:mpc_cost_general}
    \end{multline}
MPC can also enforce constraints, such as
    \begin{gather*}
        x_{t+\ell+1} = f(x_{t+\ell}, u_{t+\ell}, d_{t+\ell}) \quad \text{(System Dynamics)} \\
        u_{min} \leq u_{t+\ell} \leq u_{max} \quad \text{(Input Constraints).}
    \end{gather*}
After solving the optimization problem at time ($t$), MPC implements the first planned action  ($u^*_{t}$). At the next time step ($t+1$), the process repeats: the state is updated, the prediction horizon shifts forward, and a new optimization problem is solved based on the latest information. This receding horizon principle allows MPC to react to disturbances and model inaccuracies.

MPC is well-suited for building HVAC control due to its ability to incorporate operational constraints, optimize performance based on future predictions, and trade off competing objectives \cite{drgovna2020all}.

\subsection{Differentiable MPC Policy and Ibex-RL}
\label{sub:diff_mpc_gnurl}

While standard MPC relies on a predefined model and cost function, the Differentiable MPC policy embeds the MPC optimization within an end-to-end learning framework \cite{amos_differentiable_2019}. This approach efficiently computes gradients of the optimal control action with respect to the model and cost parameters. This is achieved via implicit differentiation through the Karush-Kuhn-Tucker (KKT) optimality conditions of the underlying MPC optimization, bypassing the need for computationally expensive backpropagation through the iterative solver \cite{amos_differentiable_2019}. Differentiable MPC enables simultaneous, gradient-based learning and adaptation of both system dynamics parameters ($\theta_{state}$) and internal quadratic cost parameters ($\theta_{cost}$). The framework also manages system constraints, potentially handling non-convexities using Projected-Newton optimization \cite{tassa2012synthesis}.

Gnu-RL \cite{chen_gnu-rl_2019} applied Differentiable MPC to HVAC using a linear state-space model for dynamics. Ibex-RL \cite{mulayim2025physics} extends this by enforcing a physics-informed structure (RC network) on the system matrices. The dynamics at time $t$ are
\begin{equation}
\begin{aligned}
x_{t+1} &= A x_t + B_u u_t + B_d d_t . \label{eq:dynamics} \\
\end{aligned}
\end{equation}
Here $A \in \mathbb{R}^{n_x \times n_x}$ is the state-to-state dynamics matrix, $B_u \in \mathbb{R}^{n_x \times n_u}$ is the control input matrix, and $B_d \in \mathbb{R}^{n_x \times n_d}$ is the disturbance input matrix. Ibex-RL \cite{mulayim2025physics} builds these matrices using parameters of a physics-informed structure that we describe later (see Eq.~\eqref{eq:modified_matrix}).

This dynamics model is coupled with with an internal quadratic cost function $J(U_t,x_t)$ within the Differentiable MPC policy:
\begin{align}
\min_{u_{t:t+L-1}} \quad & \sum_{\ell=0}^{L-1} \left( \frac{1}{2} x_{t+\ell+1}^T O x_{t+\ell+1} - \right. \nonumber \\
& \left. x_{\text{target},t+\ell+1}^T O x_{t+\ell+1} + \frac{1}{2} u_{t+\ell}^T R u_{t+\ell} \right) \label{eq:cost}  \nonumber \\
\text{subject to} \quad & x_{t+\ell+1} = f(x_{t+\ell}, u_{t+\ell}; \theta_{\text{state}}) \nonumber \\
& x_t = x_{\text{init}}  \nonumber \\ 
& \underline{u} \le u_{t+\ell} \le \overline{u}, \quad \forall \ell \in \{0, \dots, L-1\} 
\end{align}

Alternatively, this objective function can be expressed in a canonical quadratic form as:
\begin{align}
\min_{U_t} \quad & J(U_t, x_t) \nonumber \\
&= \frac{1}{2} x_t^T O_t x_t + p_t^T x_t + \frac{1}{2} u_t^T R_t u_t + s_t^T u_t \label{eq:cost_canonical} \\
&= \frac{1}{2} \underbrace{\begin{bmatrix} x_t^T & u_t^T \end{bmatrix}}_{\tau_t^T}
\underbrace{\begin{bmatrix} O_t & 0 \\ 0 & R_t \end{bmatrix}}_{C_t} \underbrace{\begin{bmatrix} x_t \\ u_t \end{bmatrix}}_{\tau_t} + \underbrace{\begin{bmatrix} p_t^T & s_t^T \end{bmatrix}}_{c_t^T} \underbrace{\begin{bmatrix} x_t \\ u_t \end{bmatrix}}_{\tau_t} \nonumber
\end{align}
Here $O$ and $R$ penalize state deviations and control effort, respectively, while $p_t$ and $s_t$ represent linear costs (i.e., related to setpoint tracking via $p_t = -O_t x_{\text{target},t}$ and $s_t=0$, respectively). 
The parameters defining the dynamics ($\theta_{state} =\{A, B_u, B_d\}$) and cost ($\theta_{cost}=\{O, R\}$) can be learned from historical data. While Differentiable MPC, in its experiments, learned this cost parameters $\theta_{cost}$ from the expert demonstration data, Gnu-RL requires the manual configuration of these parameters. Ibex-RL, on the other hand, learns them through imitation and online learning, as we detail in Section \ref{sec:objective_calibration}.

\section{Algorithmic Approach}
\label{sec:approach}
This section details the design and implementation of the MPC and model-based RL controllers tested for this study. We focus on tailoring the Ibex-RL approach \cite{mulayim2025physics} to achieve comparability with MPC while promoting safety and interpretability. Figure \ref{fig:rlandmpc} shows a high-level overview of MPC and RL architectures. Algorithms \ref{alg:mpc_control} and \ref{alg:rl_control} provide MPC and RL pseudocode.

\new{As discussed in Section \ref{sec:RLreview}, the boundary between MPC and model-based RL is blurry. The MPC variant implemented here and Ibex-RL both use the same thermal circuit model structure. However, the MPC variant identifies the model parameters through offline system identification and does not adapt them online. Ibex-RL identifies the model parameters (and cost function parameters) through offline imitation learning, then continuously adapts parameters using online learning.}

\begin{algorithm}[ht]
\caption{MPC}
\label{alg:mpc_control}
\begin{algorithmic}[1]

\Statex \textbf{Offline Phase: System Identification}
\State \textbf{Input:} Historical building data $\{x, u, d\}$.

\State Set deep mass temperature $T_m$ to the average historical indoor temperature.
\State Estimate outdoor resistance $R_{\text{out}}$ using linear regression on steady-state data.
\State Co-determine $R_m$ and $C$ via regression on unsteady data and a grid search over $R_m$.
\State $\theta_{\text{state}} \leftarrow \{T_m, R_{\text{out}}, R_m, C\}$.

\State Train an SVM model to predict exogenous heat gain $\dot{Q}_e$ from weather and time features.

\State \textbf{Output:} Identified system dynamics $f(\theta_{\text{state}}, \texttt{SVM})$.

\Statex \hrulefill
\Statex \textbf{Deployment:}
\State \textbf{Initialize:} $t \leftarrow 0$.

\State \textbf{For} each control step $t = 0, 1, 2, \dots$:
\State \quad Get current state $x_t$,  and future disturbances $d_{t:t+L}$
\State \quad \textbf{If} $t \bmod M = 0$ (every 12 hours):
\State \quad \quad Simulate system for a set of candidate weights $\{w_{c,i}\}$.
\State \quad \quad Find the minimum $w_{c,i}$ such that predicted PPD $< 10\%$.
\State \quad \quad \textbf{If} daytime:
\State \quad \quad \quad Set $w_c = 1.1 w_{c,i}$.
\State \quad \quad \textbf{Else} night:
\State \quad \quad \quad Set $w_c = 0.2 w_{c,i}$.
\State \quad Forecast exogenous heat gains using SVM.
\State \quad Solve MPC optimization:
\State \quad \quad $(U^*_t,x^*_t) \leftarrow \arg\min J(U_t, x_t)$ 
\State \quad \quad subject to constraints.
\State \quad Apply the first control action $u^*_t$ to the system.

\end{algorithmic}
\end{algorithm}

\begin{algorithm}[ht]
\caption{RL}
\label{alg:rl_control}
\begin{algorithmic}[1]

\Statex \textbf{Offline Phase: Imitation Learning}
\State \textbf{Input:} Historical building data $\{x, u, d\}$.
\State \textbf{Initialize:} Learnable parameters $\theta$.
\State \quad $\theta_{\text{state}} \leftarrow \{C, R_m, R_{\text{out}}, T_m, \eta, A_{\text{eff}}\}$
\State \quad $\theta_{\text{cost}} \leftarrow \{O, R_{hp/bh}\}$
\State Minimize imitation loss across a size $M$ batch:
\State \quad $\mathcal{L}_{\text{imit}}(\theta) = \frac{1}{M}\sum_{t=1}^{M} \|x_{t+1} - x^*_{t+1}(\theta)\|_2^2 + \lambda \|u_t - u^*_t(\theta)\|_2^2$
\State \quad \textbf{Update} $\theta$ via gradient descent: 
\State \quad $\theta \leftarrow \theta - \alpha_{\text{imit}} \nabla_{\theta} \mathcal{L}_{\text{imit}}(\theta)$

\State \textbf{Output:} State and cost parameters $\theta_{\text{init}}=\theta$

\Statex \hrulefill
\Statex \textbf{Deployment: Online Learning}
\State \textbf{Initialize:} $\theta \leftarrow \theta_{\text{init}}$, $t \leftarrow 0$
\State \textbf{For} each control step $t = 0, 1, 2, \dots$:
\State \quad Get current state $x_t$, and future disturbances $d_{t:t+L}$
\State \quad Solve differentiable MPC over horizon $L$ using $\theta$
\State \quad Compute planned state and action sequences $X_{t}^*$, $U_t^*$
\State \quad Apply the first action: $u^*_t$ 
\State \quad Observe new state $x_{t+1}$

\State \quad \textbf{If} $t \bmod M = 0$ (every midnight):
\State \quad \quad Minimize state prediction loss:
\State \quad \quad $\mathcal{L}_{\text{state}} = \frac{1}{M} \sum_{k=t-M+1}^{t} \|x_{k+1} - x^*_{k+1}\|_2^2$
\State \quad \quad \textbf{Update} $\theta_{\text{state}} \leftarrow \theta_{\text{state}} - \alpha_{\text{state}} \nabla_{\theta_{\text{state}}} \mathcal{L}_{\text{state}}$

\State \quad Estimate cumulative reward:
\State \quad $\hat{R_t} = \sum_{\ell=0}^{L-1} r(x^*_{t+\ell+1}, u^*_{t+\ell}, d_{t+\ell})$
\State \quad \textbf{Update} $\theta_{\text{cost}} \leftarrow \theta_{\text{cost}} + \alpha_{\text{cost}} \nabla_{\theta_{\text{cost}}} \hat{R}_t$

\State \quad $t \leftarrow t + 1$
\end{algorithmic}
\end{algorithm}

Gnu-RL fits generic system matrices using linear regression, which lacks direct physical meaning, hindering interpretability and direct comparison with physics-based models like those typically used in MPC. To overcome this, Ibex-RL uses the same thermal circuit model structure employed by MPC (detailed in Section \ref{sec:system_dynamics_model}) and learns the model parameters ($\theta_{state}$) from data using gradient-based methods.

The Differentiable MPC policy within the Ibex-RL framework requires an internal quadratic cost function (parameterized by $\theta_{cost}$, see Eq. \eqref{eq:cost}). This quadratic structure may not align with the potentially non-quadratic objective function (in this case, Eq. \eqref{eq:mpc_objective} optimized by MPC, as detailed in Section \ref{sec:objective_calibration}). Ibex-RL addresses this challenge through a two-stage strategy. First, the quadratic cost parameters ($\theta_{cost}$) are initialized through imitation learning based on the behavior of an existing controller using historical data (minimizing $\mathcal{L}_{\text{imit}}$, Eq. \eqref{eq:imit_loss}), aiming to provide a safe, if suboptimal, starting point for deployment. Second, during deployment, $\theta_{cost}$  is continuously adapted using policy gradients derived from maximizing a non-quadratic cumulative reward signal that has the same function structure as the MPC objective function (albeit with a different $w_c$). In parallel, Ibex-RL continually updates the model parameters $\theta_{state}$ using the state loss. This online calibration process aims to align the effective behavior of the RL agent with the true objectives, despite the differing internal cost function structures (quadratic vs. non-quadratic).

The remaining subsections describe the thermal circuit model (Section \ref{sec:system_dynamics_model}) and the objective function handling, including the online quadratic cost calibration (Section \ref{sec:objective_calibration}).

\subsection{System Dynamics Model}
\label{sec:system_dynamics_model}
\begin{figure}[ht]
    \centering
    \includegraphics[width=0.9\columnwidth]{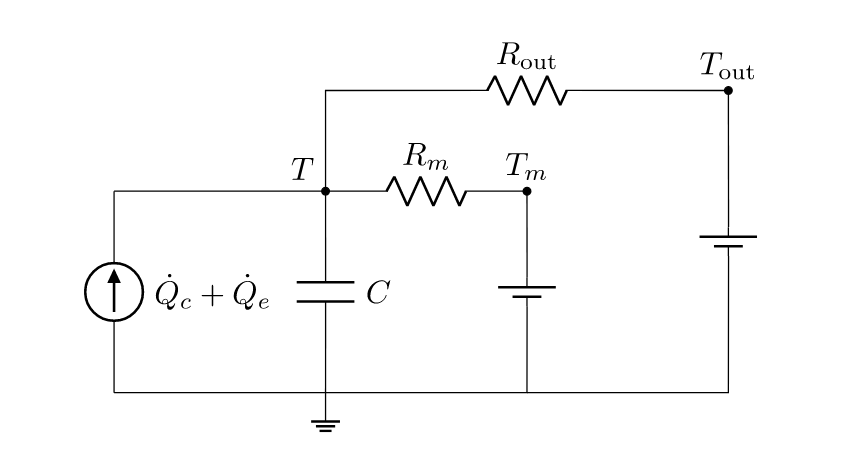}
    \caption{Thermal circuit model of the testbed.}
    \label{fig:RC}
\end{figure}

MPC uses a 2R1C thermal circuit model (shown in Figure \ref{fig:RC}) to model the system dynamics. To facilitate a direct comparison and promote interpretability, we integrated this same 2R1C architecture into Ibex-RL. \new{This low-order structure was selected for three reasons: (1) maintaining model parity between controllers ensures that observed performance differences are attributable to the control algorithm rather than model fidelity; (2) for lightweight wood-frame residential buildings, prior work has demonstrated that low-order models avoid overfitting while adequately capturing dominant thermal dynamics for control purposes \cite{blum2019practical}; and (3) a low-order model reduces data requirements and computational cost, supporting the scalability objectives of this work.}

The continuous-time 2R1C dynamics are:
\begin{multline}
\label{eq:generic_21rc}
C \frac{dT}{dt} = \frac{T_{m} - T}{R_{m}} + \frac{T_{out} - T}{R_{out}} + \dot{Q}_c + \dot{Q}_e ,
\end{multline}
where $T$ is the indoor air temperature (measured in the return duct), $T_{out}$ is the outdoor air temperature, and $T_m$ is the thermal mass temperature. \new{The model parameters include the thermal resistances between the indoor air and the thermal mass ($R_m$) and between the indoor and outdoor air ($R_{out}$), along with the overall thermal capacitance ($C$). The thermal resistances represent the combined impedance to convective, conductive, and (linearized) radiative heat transfer.} The term $\dot{Q}_c$ is the controllable thermal power input from the HVAC system (summing contributions from the heat pump and backup heat), while $\dot{Q}_e$ represents exogenous heat gains from sunlight, body heat, plug loads, etc.

While MPC and Ibex-RL use the same 2R1C model structure, they use different methods to identify the model parameters. The MPC parameters were identified through a multi-step process relying on specific conditions within historical data. This involved assuming a constant deep mass temperature ($T_m$) set to the average indoor temperature observed during the training period (a simplification deemed acceptable for the forced-air system). The outdoor resistance ($R_{out}$) was then estimated via linear regression using steady nighttime data. Subsequently, the internal resistance ($R_m$) and capacitance ($C$) were co-determined using linear regression on unsteady data, combined with a grid search over $R_m$ for validation. The exogenous thermal power ($\dot{Q}_e$) for the MPC model was predicted using a separate supervised learning model – specifically, a Support Vector Machine (SVM) trained on outdoor temperature, solar irradiance, wind speed, and time features. This SVM provided a data-driven forecast integrating solar gains and other unmodeled thermal loads. \new{Alternatively, the full dataset could be used to jointly fit all model parameters by minimizing a global prediction error. Ibex-RL essentially implements this approach.}

Ibex-RL was designed for autonomous, end-to-end parameter fitting, avoiding the need for specialized data subsets (like nighttime-only) or auxiliary predictive models (like the SVM for $\dot{Q}_e$). To achieve this, Ibex-RL's 2R1C formulation models the exogenous heat gain using an effective solar area ($A_{eff}$) as a learnable parameter:
\begin{multline}
\label{eq:thermal_balance}
C \frac{dT}{dt} = \frac{T_{m} - T}{R_{m}} + \frac{T_{out} - T}{R_{out}} \\
+ \underbrace{COP\bigl(T_{out}\bigr) P_{HP} + \eta P_{BH}}_{\dot{Q}_c} + \underbrace{A_{eff} I_{sol}}_{\dot{Q}_e} .
\end{multline}
Here $I_{sol}$ (kW/m$^2$) is the global solar irradiance on a horizontal surface, $A_{eff}$ is the effective area for solar heat gain, $P_{HP}$ and $P_{BH}$ are the electric power inputs for the heat pump and backup heat, and $\eta$ is the backup heat efficiency (assumed 1 by MPC). MPC and Ibex-RL use the same $COP(T_{out})$ function.

\new{Both MPC and Ibex-RL treat the internal state space (indoor temperature) and action space (HVAC power inputs) as continuous; no discretization is applied within the policy or learning mechanisms. However, the thermostat can only accept temperature setpoints in half-degree (Celsius) increments. Therefore, a post-processing layer rounds the target temperature outputs from MPC and Ibex-RL ($x^*_{t+1}$) to the nearest 0.5 $^{\circ}$C.}

The continuous-time dynamics \eqref{eq:thermal_balance} are mapped to a discrete-time state-space representation. This is achieved by first defining the continuous-time matrices based on the physical parameters, and then converting them to a discrete-time model using a zero-order hold discretization for a sampling period \(\Delta t\).
The continuous-time system matrix \(A_c\), input matrix \(B_{uc}\), and disturbance matrix \(B_{dc}\) are:
\begin{equation}
\label{eq:modified_matrix}
\begin{split}
A_c &= - \left( \frac{1}{R_mC} + \frac{1}{R_{out}C} \right) \\
B_{uc}(T_{out,t}) &= \begin{bmatrix} \frac{\text{COP}(T_{out,t})}{C} & \frac{\eta}{C} \end{bmatrix} \\
B_{dc} &= \begin{bmatrix} \frac{1}{R_mC} & \frac{1}{R_{out}C} & \frac{A_{eff}}{C} \end{bmatrix} .
\end{split}
\end{equation}
Using a standard zero-order hold, these matrices are discretized to form 
\begin{align*}
A &= e^{A_c \Delta t} \\
B_u[T_{out,t}] &= A_c^{-1} (A - I) B_{uc}(T_{out,t}) \\
B_d &= A_c^{-1} (A - I) B_{dc}
\end{align*}
in the discrete-time dynamics, \(x_{t+1} = A x_t + B_u[T_{out,t}] u_t + B_d d_t\). The state vector \(x_t\) contains the indoor temperature \(T_t\). The disturbance vector \(d_t\) includes the thermal mass temperature \(T_m\) (treated as a measurement), the outdoor temperature \(T_{out, t}\), and the solar irradiance \(I_{sol, t}\). The control inputs \(u_t\) are the heat pump and backup heat powers \(P_{hp, t}\) and \(P_{bh, t}\).

For the RL system dynamics, gradients are computed during the imitation learning phase to fit the physical parameters \(\theta_{state} = \{C, R_m, R_{out}, T_m, \eta, A_{eff}\}\). These parameters, in turn, define the state-space matrices ($A, B_u, B_d$), where the control matrix $B_u$ is recalculated at each timestep to account for the time-varying heat pump Coefficient of Performance (COP). Simultaneously, the framework learns parameters for the internal cost function \eqref{eq:cost}, \(\theta_{cost} = \{O_t, R_t\}\), where \(O_t \in \mathbb{R}\) is the state cost weight and \(R_t \in \mathbb{R}^{2 \times 1}\) contains the control cost weights for \(P_{HP}\) and \(P_{BH}\).

All parameters $\theta = \{\theta_{state}, \theta_{cost}\}$ are jointly learned to minimize the imitation loss function:
\begin{equation}
\mathcal{L}_{\text{imit}}(\theta) =\frac{1}{M} \sum_{t}^M \|x_{t+1} - x^*_{t+1}(\theta) \|_2^2 + \lambda \|u_t - u^*_t(\theta)\|_2^2
\label{eq:imit_loss}
\end{equation}
where $x$ and $u$ are the state and action from a batch of $M$ expert demonstrations, while $x^*(\theta)$ and $u^*(\theta)$ are the predicted next state and action generated by the policy parameterized by $\theta$. The loss penalizes deviations in both predicted next states ($\mathcal{L}_{\text{state}}=\frac{1}{M} \sum_{t}^M \|x_{t+1} - x^*_{t+1} \|_2^2$) and chosen control actions ($\mathcal{L}_{\text{action}}=\frac{1}{M} \sum_{t}^M \|u_t - u^*_t\|_2^2$), balanced by the weight $\lambda$. \new{Prior work on Ibex-RL \cite{mulayim2025physics} and Gnu-RL \cite{chen_gnu-rl_2019} showed that fixing the control inputs and minimizing the state prediction error yields a lower state prediction error than jointly minimizing the state and action prediction errors, but results in worse controller performance. This disconnect between predictive accuracy and control performance is a challenge addressed in the field of decision-focused learning \cite{hai2025decision}.}

In summary, while the MPC approach fits four physical parameters ($C, R_m, R_{out}, T_m$) and trained an SVM to predict exogenous heat gains, Ibex-RL jointly fits six physical parameters and three cost parameters offline through imitation learning. Ibex-RL then adapts the parameters using online learning, with separate gradient calculations for $\theta_{state}$ and $\theta_{cost}$.

\new{Model-free RL typically uses discrete state and action spaces and iterates toward an optimal policy, generally without convergence guarantees for finite sample sizes. The model-based Ibex-RL approach, by contrast, uses continuous state and action spaces, similar to most MPC variants. The Differentiable MPC policy underlying Ibex-RL solves a convex quadratic program to global optimality at each time step.}

Finally, the RL policy, through Differentiable MPC policy, allows the direct enforcement of box constraints on control inputs, identical to those used by MPC: $P_{HP}^{\text{min}} \leq P_{HP} \leq P_{HP}^{\text{max}}$ and $P_{BH}^{\text{min}} \leq P_{BH} \leq P_{BH}^{\text{max}}$.

\subsection{Objective Function and Quadratic Cost Calibration}
\label{sec:objective_calibration}

In the MPC cost function,
\begin{multline}
  w_d \max(u^*_t, \dots, u^*_{t+L-1}) + \Delta t \sum_{\ell=0}^{L-1} \left[ w_{e} u^*_{t+\ell} \right. \\ 
  + \left. w_{c,t+\ell+1} \left| x^*_{t+\ell+1} - x_{target,t+\ell+1} \right| \right] ,
    \label{eq:mpc_objective}
\end{multline}
$w_d$ (\$/kW) is the peak demand price (set to \$0.8/kW)\footnote{\new{The $w_d$ term penalizes peak demand over the one-day horizon as an approximate stand-in for the peak demand over a monthly billing period. The price $w_d$ was scaled down from a typical monthly price to reflect the one-day horizon \cite{pergantis_field_2024}.}}, $w_e$ (\$/kWh) is the electrical energy price (set to \$0.15/kWh, based on local rates), and $w_c$ (\$/($^{\circ}$C$\cdot$h)) is the thermal discomfort price. \new{To balance trade-offs between energy costs and discomfort, MPC adapted $w_c$ online. Every 12 hours, an optimization loop swept candidate $w_c$ values, selecting the lowest one predicted to maintain the time-average Predicted Percentage of Dissatisfied (PPD, a measure of thermal discomfort) below 10\%. This selected $w_c$ was then scaled by hand-tuned factors of 1.1 during the day and 0.2 overnight \cite{pergantis_field_2024}. This tuning process aimed to strike a reasonable balance between energy savings and occupant comfort in all hours of the day. The lower nighttime scaling factor (0.2) reflects the occupant-selected setpoint schedule, which allows lower temperatures overnight (18$\,^\circ$C vs.\ 20$\,^\circ$C during the day). A high nighttime discomfort penalty would prevent the controller from pre-heating in anticipation of the morning recovery period.}

Following initialization via imitation learning, Ibex-RL dynamically updates the internal cost parameters, $\theta_{cost}$. This adaptation uses gradients derived from maximizing a non-quadratic reward signal $r_t$ that mirrors the structure of the MPC cost function. The objective is to adjust $\theta_{cost}$ such that the resulting control policy achieves the true control objectives of reducing peak demand, energy costs, and deviations of the indoor temperature from the occupant preference. The calibration process, detailed in Section \ref{sec:objective_calibration}, evaluates the planned states and actions ($x^*_{t+1}, \dots, x^*_{t+L}$ and $u^*_t, \dots, u^*_{t+L-1}$) over the prediction horizon. 

\new{While MPC dynamically updated the discomfort price $w_c$ online to approximately maintain the time-average PPD below 10\%, this was impractical in Ibex-RL. Within the reward function, $w_c$ only influences the gradient used to update the internal quadratic cost parameters $\theta_{cost}$. It does not directly alter the planned actions or states in a way that allows PPD evaluation for each candidate $w_c$ value during a sweep. Ibex-RL therefore uses a static value for $w_c$ (set to 3\,\$/($^{\circ}$C$\cdot$h)). This simplification impacts the comfort-cost trade-off, a factor to consider when interpreting the results. The value of 3\,\$/($^\circ$C$\cdot$h) was hand-tuned based on preliminary offline simulations to produce a reasonable comfort-energy balance. While no formal sensitivity study was conducted on this parameter during deployment, the online adaptation of the cost parameters $\theta_{cost}$ --- particularly the monotonic increase in the state penalty $O$ shown in Figure \ref{fig:parameters} --- partially compensates for the static $w_c$ by increasingly emphasizing comfort over time.}

\section{Experiments}
\label{sec:experiments}

This section details the experimental setup for field deployments in the test house (Figure \ref{fig:dc_house}). We describe the datasets used for controller training, validation, and evaluation, outline the pre-training procedure for MPC and RL via imitation learning, and specify the common input data, decision variables, and operational settings applied across the controllers during the comparative tests. 

\subsection{Input Data and Decision Variables}
In addition to the controller-specific parameters for system dynamics and objectives, several common inputs and settings were used for both MPC and RL during deployment. 
\new{MPC and RL both used measurements of the indoor temperature ($T$) and the electric power inputs to the heat pump ($P_{HP}$) and auxiliary heating elements ($P_{BH}$). MPC and RL both actuated the indoor temperature setpoint, which they sent to the thermostat via API.} The indoor temperature was measured using a sensor located in the return air duct to capture a representative measurement of the mixed-air temperature. Forecasts for external conditions, specifically outdoor temperature ($T_{out}$) and global solar irradiance ($I_{sol}$), were obtained via the Okiolab\footnote{\url{https://oikolab.com/}} API. Both controllers operated with a discrete time step of $\Delta t = 1\,$hour and used a prediction horizon of $L=24$ hours. Finally, a common user-defined temperature setpoint (shown as $x_{target,t}$ in Eq. \ref{eq:mpc_objective}) schedule was implemented for both systems: $18\,^{\circ}$C from midnight to 6:00 AM and $20\,^{\circ}$C during all other hours.

One practical challenge was translating the planned MPC and RL actions into commands for the physical HVAC system. While both MPC and RL internally planned HVAC power and indoor temperature trajectories, the heat pump thermostat's API only accepted indoor temperature setpoints. Therefore, an indirect actuation strategy was implemented for both controllers: The thermostat setpoint for the upcoming control interval was set to the next planned state temperature, $x^*_{t+1}$, rounded to the nearest 0.5 $^\circ$C. This approach implicitly assumes that the heat pump manufacturer's device-level control system accurately tracks indoor temperature setpoints.

\subsection{Data}
This study uses three kinds of datasets for model development and analysis. Training data was used to fit the core models: the system dynamics for MPC, and the state and cost parameters via imitation loss for RL. Validation data then served to test the predictive accuracy of these fitted models. Finally, evaluation data comprises the operational data collected during the deployment of each controller to analyze their real-world performance.

\subsubsection{MPC and RL Training and Validation Data}
\label{sec:train_data}

\begin{figure}[ht]
    \centering
    \includegraphics[width=0.8\columnwidth]{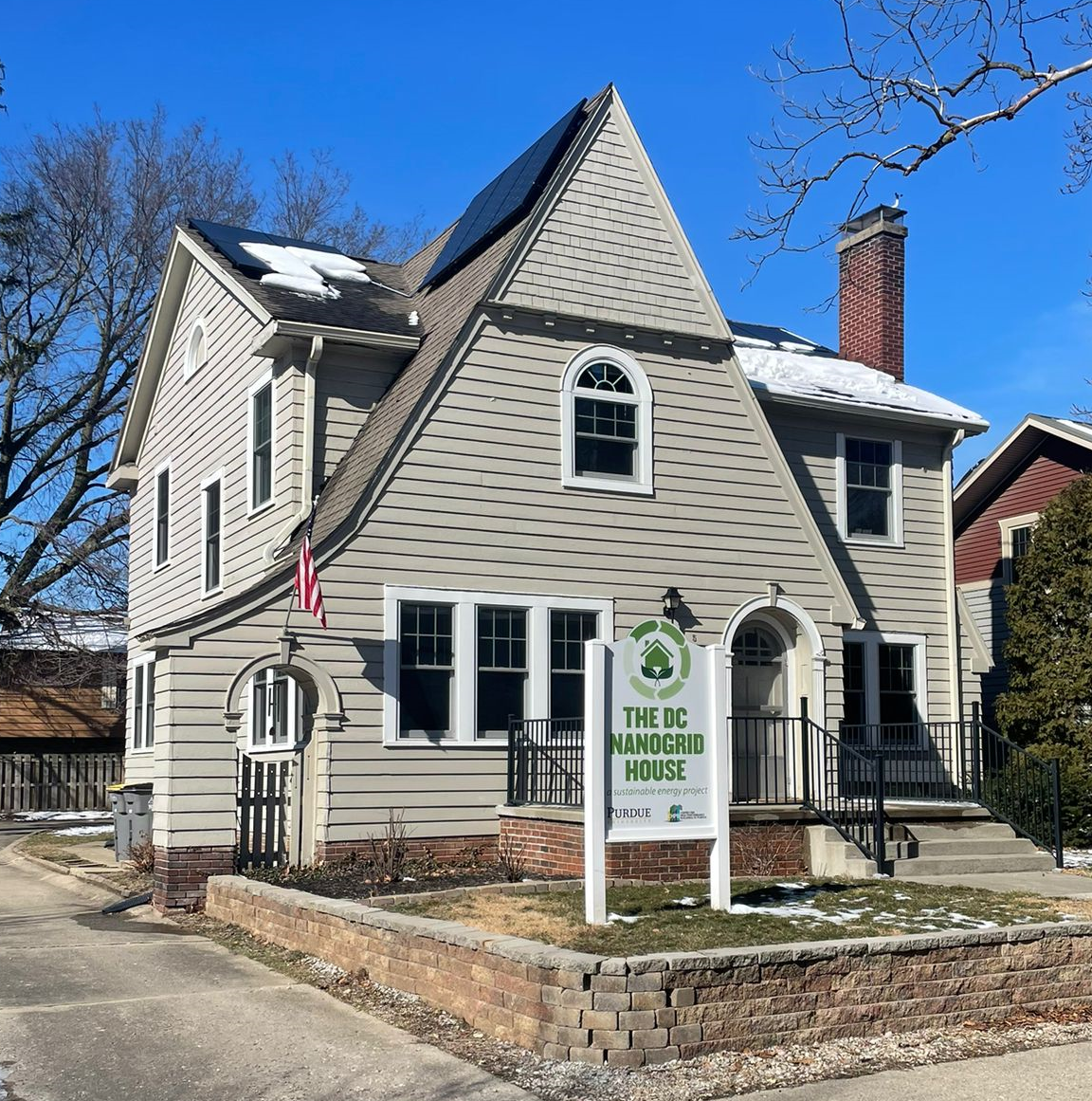}
    \caption{The test house is a 208 m$^\text{2}$, 1920s-era detached house with all-electric appliances in West Lafayette, Indiana. }
    \label{fig:dc_house}
\end{figure}

\new{For model training and validation, MPC and Ibex-RL used power and temperature data from the test house (Figure \ref{fig:dc_house}). The five-minute data were resampled to hourly intervals to coincide with hourly historical weather data from Okiolab. In the training and validation periods, the heat pump was operated by the manufacturer's device-level control system, with no supervisory control.}

\new{MPC's model identification used training data from November 11 to December 10, 2022, and validation data from December 10–29, 2022. Ibex-RL's imitation learning used training data from November 1–29, 2023, and validation data from December 15–30, 2023. The training and validation dataset sizes were comparable for MPC and RL, although collected during different time periods.}

\new{The MPC training data included night setbacks to the thermostat temperature setpoint, programmed by occupants. Each night setback provided a downward step response in the evening and an upward step response in the morning. While we did not formally quantify information content or persistence of excitation in the training data, we found the passive excitation from night setbacks to be sufficient for MPC parameter identification. The Ibex-RL training data, by contrast, contained mostly constant setpoints. This likely made parameter identification more challenging for Ibex-RL than for MPC due to a lack of indoor temperature variability, a known challenge for identifying building thermal models \cite{blum2019practical, humidity_dchouse, WANG2023113026}. }

\subsubsection{Baseline, MPC, and RL Evaluation Data}
\label{sec:evaluation_data}

We benchmark MPC and Ibex-RL performance against a baseline without supervisory control. With or without supervisory control, the heat pump manufacturer's control system adjusted device-level variables, such as compressor and fan speeds, to track the thermostat temperature setpoint. The supervisory MPC and Ibex-RL control systems sent temperature setpoint adjustments to the thermostat. Without supervisory control, the setpoint followed occupant preferences. 

For the baseline and MPC, data was initially collected between November of 2022 and April of 2023. During this period, baseline and MPC operation were interleaved. After removing days with missing values, the effective (but not continuous) period for baseline data was December 11, 2022, through April 4, 2023, and for MPC data was February 1 through March 30, 2023. To facilitate analysis of daily energy data, we filtered these datasets, retaining only days where a single controller operated for 20 hours or more. This resulted in 65 days of baseline data and 23 days of MPC data.

Ibex-RL was deployed from January 23 to February 23, 2025. For performance analysis, we applied the same filtering criterion, excluding days with less than 20 hours of operation or those affected by communication issues. This yielded 23 days of RL data.

\subsection{Pre-training}

MPC training for the system dynamics model following Section \ref{sec:system_dynamics_model}, resulted in an average $\mathcal{L}_{\text{state}}= 0.41$°C. Its parameters were $R_m = 1.06$ °C/kW, $R_{\text{out}} = 2.04$ °C/kW, $C = 6.5$ kWh/°C, and $T_m = 20.6$ °C. 

Ibex-RL training involves two components that are learned simultaneously: (a) learning the system dynamics, represented by the 2R1C thermal model in Eq. \eqref{eq:thermal_balance}, and (b) learning the desired control behavior, represented by the cost function in Eq. \eqref{eq:cost}. We tested various hyperparameters using validation set to optimize both state and action losses. 

\begin{figure}[!t]
    \centering
    \includegraphics[width=1\columnwidth]{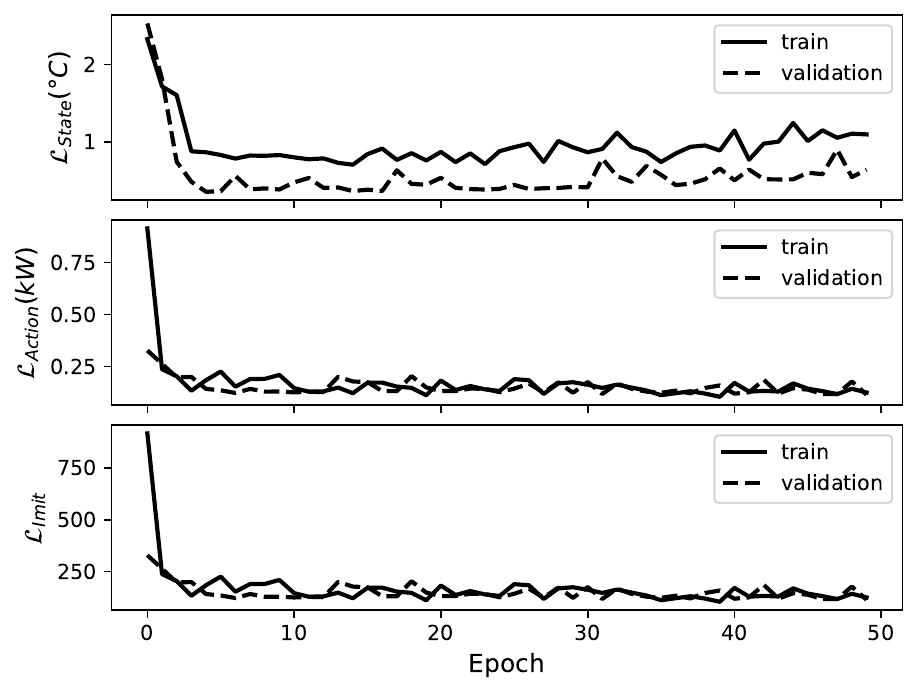}
    \caption{State and action losses for Ibex-RL imitation learning with the hyperparameter combination $\alpha_{\text{imit}}=0.05$ and $\lambda=1000$, which gave the lowest action loss.}
    \label{fig:imit_learning}
\end{figure}

During training in imitation learning (using data explained in Section \ref{sec:train_data}), there are two main hyperparameters: The learning rate ($\alpha_{\text{imit}}$) and the weight for balancing the relative importance of actions and next-state predictions ($\lambda$). We trained the model with varying values of $\alpha_{\text{imit}} \in \{0.05, 0.005, 0.0005\}$ and $\lambda \in \{1, 1000\}$ for 50 epochs with a batch size of $M=24$. Each epoch was run for 30 days of hourly data where instances were sampled randomly. The results shown in Figure \ref{fig:imit_learning} correspond to the combination $\{0.05, 1000\}$, which was selected because it produced the lowest $\mathcal{L}_{\text{action}}$ (0.11 kW) based on the dataset. 

With this hyperparameter combination, the validation loss was $\mathcal{L}_{\text{state}} = 0.64$°C. While other combinations resulted in smaller $\mathcal{L}_{\text{state}}$ values, they appeared to significantly underestimate the effect of the heating system, which is checked by simulating a one hour transition with full on heating and then observing the change in predicted temperature using the system dynamics model fitted. Considering this observation and the fact that matching the behavior of the existing controller is important for promoting safety during initial deployment, we selected the aforementioned hyperparameter set.

Compared to MPC, Ibex-RL recovered similar values for the parameters common to both methods: $R_m = 1.07$ °C/kW, $R_{\text{out}} = 1.07$ °C/kW, $C = 4.8$ kWh/°C, and $T_m = 26.25 $°C. The MPC implementation used a SVM to model the exogenous heat gains, and therefore did not include the effective solar area $A_{eff}$ used in Ibex-RL. \new{Because there are no ground-truth values for these parameters, comparing the Ibex-RL values to the MPC values only gives a relative, not absolute, performance assessment. However, the similarity of the values across training methods builds confidence that both methods performed reasonably.} As models used for control do not necessarily need to mirror the environment perfectly, as long as they guide the controller in sensible directions \cite{liu2006experimental1}, we believe (and our results demonstrate) that the model fits in this section can yield performance gains.

\section{Field Deployment Results}
\label{sec:results}

This section presents the results from the month-long field deployments of MPC and Ibex-RL. We first analyze the online learning behavior and adaptation characteristics of the RL agent. We then compare MPC, Ibex-RL, and baseline performance on representative days under varying weather conditions. Finally, we compare energy savings, energy efficiency, occupant comfort, and estimated deployment labor.

\subsection{RL Controller Adaptation and Behavior}\label{sec:rl_adaptation}

\begin{figure}[h!]
\centering\includegraphics[width=\linewidth]{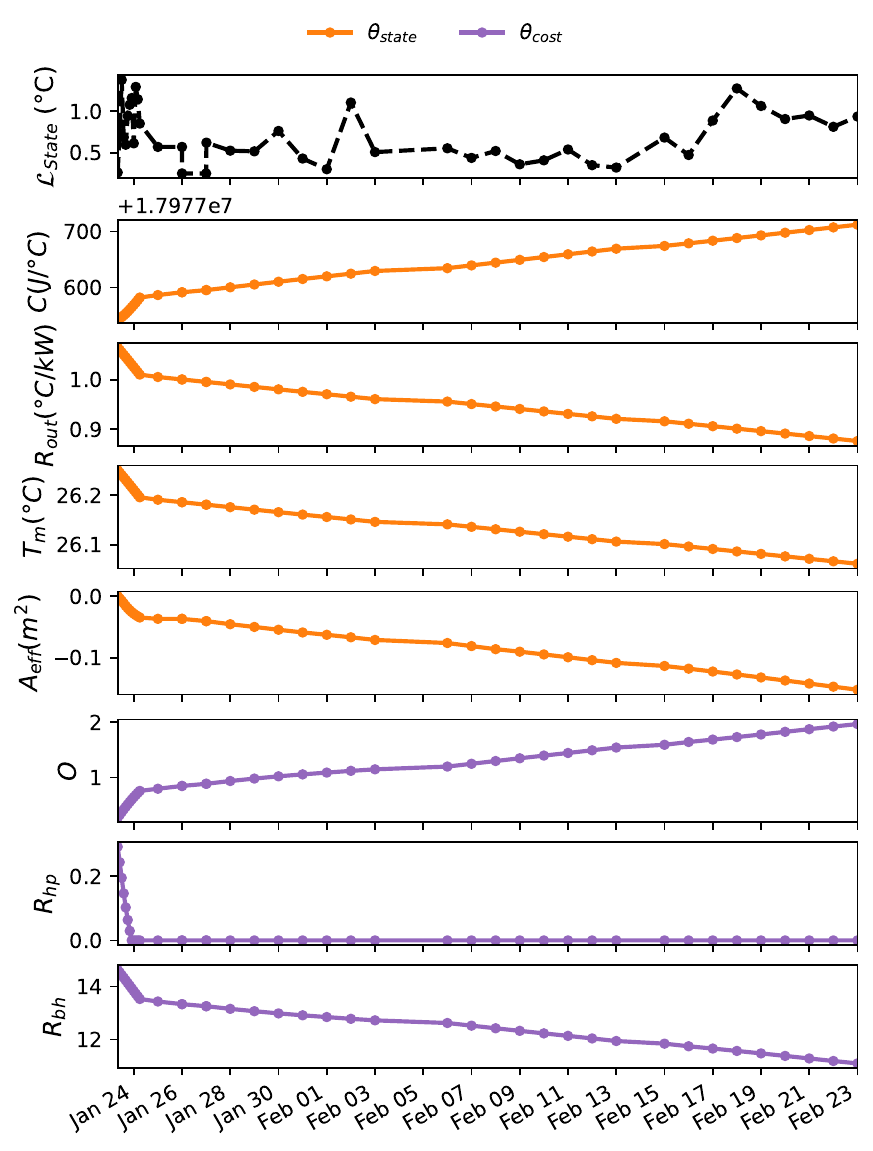} 
\caption{Evolution of the learned dynamics parameters ($\theta_{state}$) and cost weights ($\theta_{cost}$).} 
\label{fig:parameters}
\end{figure}

Ibex-RL used online learning to update both its dynamics model parameters ($\theta_{state}$) and its internal cost function parameters ($\theta_{cost}$) every 24 hours (with the exception of the first day, where updates occurred every two hours to accelerate initial adaptation) based on the latest interaction data. \new{Figure \ref{fig:parameters} shows the evolution of the learned parameters. Ibex-RL did not run long enough for most dynamics parameters (orange) or cost parameters (purple) to converge to steady values. Most parameter values were still evolving when the experiment ended on February 23. The exceptions are the heat pump cost weight $R_\text{hp}$, which converged quickly to zero, and the thermal mass temperature $T_m$, which changed by 0.15 $^\circ$C over the test month.}

The second through fifth plots in Figure \ref{fig:parameters} show the dynamics parameters. The thermal capacitance $C$ increased over time, moving closer to the value identified by MPC, though never strictly matching it. The thermal resistance to the outdoors $R_{out}$ decreased to approximately $0.9 ^{\circ}$C/kW, changing by around 0.2 $^\circ$C/kW over the test month. The thermal mass temperature $T_m$ began at $26.25^{\circ}\mathrm{C}$ (significantly higher than the MPC value of $20.6^{\circ}\mathrm{C}$) and decreased slightly to $26.1^{\circ}\mathrm{C}$. The effective solar area $A_{eff}$, which was initialized near zero during imitation learning, drifted into negative values. This physically unrealistic result suggests the model failed to capture the impact of solar gains, incorrectly implying a cooling effect from solar radiation.

\new{The bottom three plots in Figure \ref{fig:parameters} show the cost parameter evolution. The temperature weight $O$ quickly increased from about zero to one on the first day, then increased more gradually. The heat pump weight ($R_{\text{hp}}$) dropped quickly to a steady value of zero, while the backup heat weight ($R_{\text{bh}}$) decreased more gradually, remaining above 10 kW/$^\circ$C, and did not reach a steady value within the experiment duration. The combined increase in the temperature weight and decrease in the power weights can be interpreted as Ibex-RL learning to prioritize comfort more over time, and especially on the first day, when Ibex-RL kept the house quite cold. The backup heat weight remaining fairly high aligns makes sense, as backup heat is much less efficient than the heat pump. However, we do not have a clear explanation for the heat pump weight converging to zero. 
}

\new{Overall, the online learning results did not align well with our intuition. Two factors may have contributed to parameters diverging from plausible values. First, the system was not actively excited during deployment to gather information-rich data, as our goal was to compare performance under typical operating conditions similar to MPC deployment. In other words, this Ibex-RL implementation did not balance trade-offs between exploration (gathering information-rich data to improve parameter estimates) and exploitation (taking actions to maximize expected reward under the current parameter estimates), instead doing exploitation only.} 

Second, a discrepancy existed between the controller's intended action and its actual implementation. The agent planned power levels $P_{HP}, P_{BH}$ but the system was controlled by translating this into a thermostat setpoint based on the next planned state ($x^*_{t+1}$), sometimes resulting in significantly different power levels than planned. A further mismatch arose because the model used the return air temperature as its state measurement, while the thermostat used its own sensor in a different location for device-level control. Both discrepancies introduced noise into the online learning data. 

\new{Despite the online learning results not matching our intuition, Ibex-RL still saved significant energy and gave reasonable closed-loop behavior, as the remainder of this section discusses.}

\subsection{Representative Days}

\begin{figure}[ht!]
    \centering
    \includegraphics[width=\columnwidth]{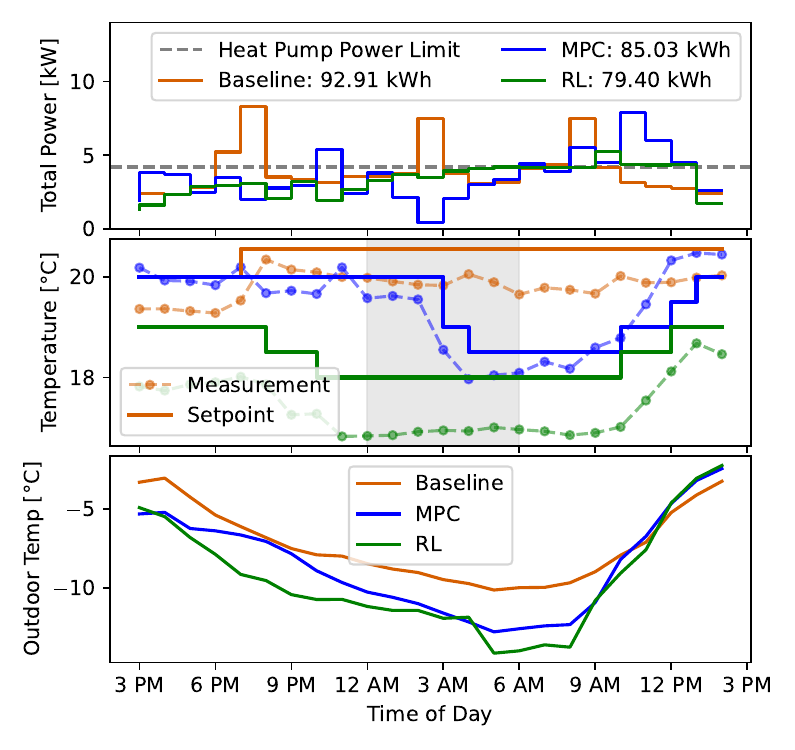}
    \caption{Cold-weather comparison. RL saves energy by keeping the house cooler and using less backup heat. MPC saves less energy but more closely tracks occupant preferences (18 $^\circ$C from midnight to 6 AM, 20 $^\circ$C otherwise). 
    }
    \label{fig:cold_day}
\end{figure}

\begin{figure}[t!]
    \centering
    \includegraphics[width=\columnwidth]{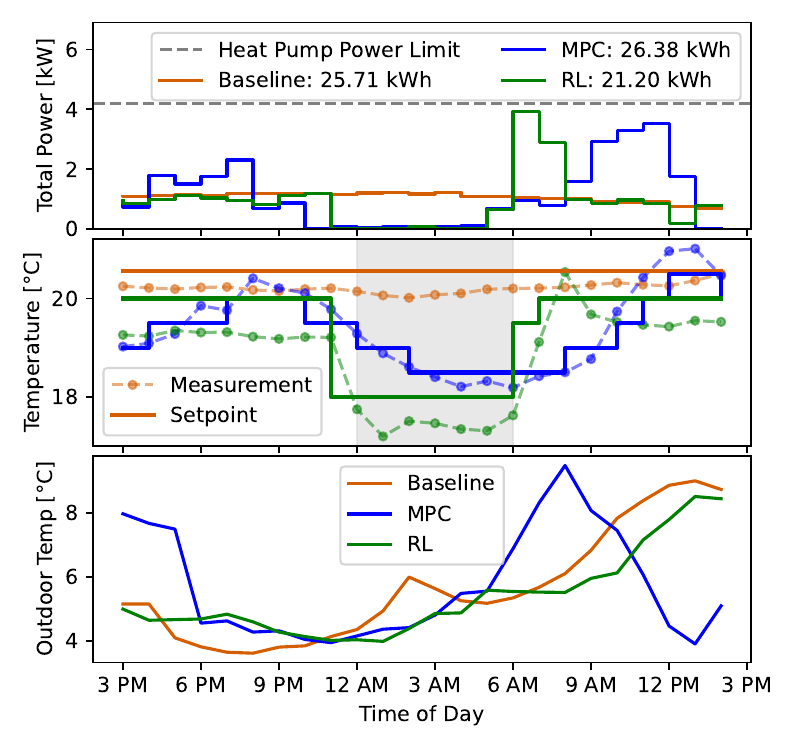}
    \caption{Warm-day comparison. RL anticipates changes in the occupant preference (20 to 18 $^\circ$C at midnight, 18 to 20 $^\circ$C at 6 AM, shown in gray), while MPC shows more gradual transitions. 
    }
    \label{fig:warm_day}
\end{figure}

We illustrate controller performance on two typical days representing different weather conditions, presented in Figures \ref{fig:cold_day} and \ref{fig:warm_day}. 
Each figure follows the same format. The top panel shows the hourly electric power inputs to the heat pump and backup heat, with total daily energy consumption for each method shown in the legend. The middle panel shows indoor temperature setpoints (dashed lines) from each controller alongside their measured return air temperatures (solid lines). The bottom panel shows the outdoor temperature profile for each controller. We visually matched outdoor temperature profiles across controllers to approximate normalize for weather variation.

Figure \ref{fig:cold_day} shows a cold-weather comparison where all controllers experienced similar outdoor temperatures, with slightly colder conditions for RL. RL generally keeps the house one to two $^\circ$C below the user preferences of 18 $^\circ$C from midnight to 6 AM and 20 $^\circ$C during all other hours, achieving the lowest energy consumption at 79.4 kWh. This behavior stems from its optimization objective that prioritizes energy savings over comfort during extreme cold conditions through the weighting parameter ($w_c$). MPC maintains setpoints closer to user preferences, resulting in higher energy use (85 kWh) but better comfort. Both approaches can configure these trade-offs through adjustable parameters like $w_c$. The baseline, constant-setpoint control uses more 92.9 kWh, activating backup heating three times compared to a single activation each under MPC and RL.

Figure \ref{fig:warm_day} shows performance under warmer conditions. RL reduces the setpoint one hour before the midnight step-change in the occupant preference, achieving the desired 18 $^\circ$C at the desired time. MPC reduces the setpoint more gradually, resulting in temporary comfort deviations. During morning warm-up periods, RL's two-stage setpoint increase (18°C to 19.5°C at 6 AM) stays within the heat pump's capacity, avoiding use of backup heat. MPC's slower response leads to lower-than-setpoint conditions in the early morning hours, although this is partially attributable to the steep outdoor temperature decline between 8 AM and 1 PM. \new{We hypothesize that these distinct variations in performance under different weather conditions are ultimately driven by the controllers' differing training data pipelines and optimization objectives.}

\subsection{Aggregate Energy Savings}

\new{We compare daily electrical energy use across the control configurations in two ways. In Section \ref{sec:savings1}, we normalize electricity use by the outdoor temperature directly. In Section \ref{sec:savings2}, we normalize electricity use by the difference between the indoor and outdoor temperatures. The first comparison includes energy savings both from reducing indoor temperatures and from operating the heat pump and backup resistive heater more efficiently (i.e., using less electricity to maintain a given indoor-outdoor temperature difference). The second comparison does not include energy savings from reducing indoor temperatures, which may also reduce occupant comfort.}

\new{We construct both energy comparisons such that the electricity savings estimate is weather-independent. We model the daily electrical energy use $E$ as affine in a temperature feature $x$:
\begin{equation}
E \approx \beta ( x - x_0 ) . \label{energyModel}
\end{equation}
In Section \ref{sec:savings1}, $x = \overline T_{out}$, the daily average outdoor temperature. In Section \ref{sec:savings2}, $x = \overline T_{out} - \overline T_{in}$, the daily average temperature difference. For the comparison in each section, we model the daily electricity savings for MPC (subscript $m$) relative to the baseline (subscript $b$) as
\begin{equation}
1 - \frac{E_m}{E_b} \approx 1 - \frac{\beta_m (x - x_0)}{\beta_b (x - x_0)} = 1 - \frac{\beta_m}{\beta_b} . \label{savingsModel}
\end{equation}
The weather-dependent $x - x_0$ terms cancel because we use the same $x$-intercept $x_0$ for all three control configurations. We model the relative daily electricity savings for RL (subscript $r$) similarly: $1 - \beta_r / \beta_b$.}

\new{In each section, we fit the slopes $\beta_b$, $\beta_m$, and $\beta_r$ and the $x$-intercept $x_0$ through a combination of linear regression and a grid search. We first specify a finite set of candidate $x_0$ values. We then use linear regression to fit the slope $\beta$ in the model \eqref{energyModel} for each $x_0$ value and each control configuration (baseline, MPC, and RL). We select the $x_0$ value that maximizes (over candidate $x_0$ values) the minimum  $R^2$ statistic over control configurations. In each section, this procedure outputs a single estimated $x$-intercept $\hat x_0$, shared across all three control configurations, as well as the three slope estimates $\hat \beta_b$, $\hat \beta_m$, and $\hat \beta_r$. The $x$-intercept and slopes are different in general for the comparisons in Section \ref{sec:savings1} vs. Section \ref{sec:savings2}, as can be seen from Figures \ref{fig:outdoor} and \ref{fig:delta}. Under the standard assumptions of linear regression, the slopes $\beta_b$, $\beta_m$, and $\beta_r$ are Gaussian. The ratios $\beta_m / \beta_b$ and $\beta_r / \beta_b$ in the savings formulas are non-Gaussian, however. We therefore estimate the savings means and confidence intervals through Monte Carlo simulation with a sample size of $10^7$.}

\subsubsection{Outdoor Temperature Fits}
\label{sec:savings1}

\begin{figure}[ht]
    \centering
    \includegraphics[width=\columnwidth]{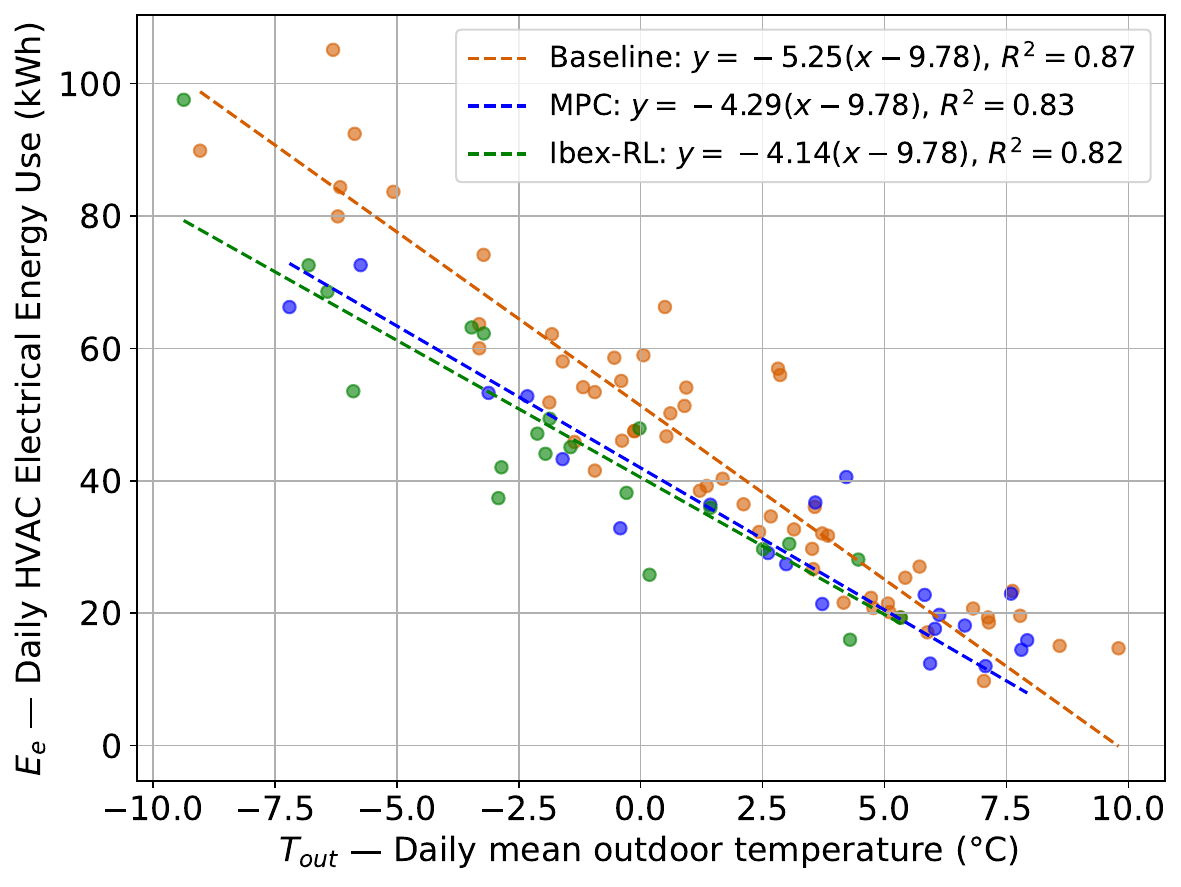}
    \caption{Fits of daily energy use vs. outdoor temperature.}
    \label{fig:outdoor}
\end{figure}

\new{Figure \ref{fig:outdoor} shows the fits from Eq. \eqref{energyModel} with $x = \overline{T}_{out}$, the daily average outdoor temperature. Each dot is the electrical energy used on a day with a particular average outdoor temperature. The dashed lines are the linear fits for the baseline (no supervisory control, orange), MPC (blue), and RL (green). All control configurations have $R^2 \ge 0.82$. RL has the least negative slope ($\hat \beta_r = -4.14$ kWh/$^\circ$C), indicating marginally lower daily electricity use than MPC ($\hat \beta_m = -4.29$ kWh/$^\circ$C) and significantly lower than the baseline ($\hat \beta_b = -5.25$ kWh/$^\circ$C).  Differences in electricity use are larger at lower outdoor temperatures.}

\new{The top half of Table \ref{tab:controller_performance} summarizes the estimated savings means and 95\% confidence intervals from Eq. \eqref{savingsModel}, alongside three hypothesis tests: (1) MPC uses less electricity than the baseline ($\beta_m < \beta_b$), (2) RL uses less electricity than the baseline ($\beta_r < \beta_b$), and (3) RL uses less electricity than MPC ($\beta_r < \beta_m$). The estimated mean relative savings for RL are 20.94\% (95\% CI: 2.60\% to 38.38\%) over the baseline, which is statistically significant ($p = 0.013, d = 2.30$). The estimated mean relative savings for MPC are 18.13\% (95\% CI: 4.38\% to 30.90\%), which is also statistically significant ($p = 0.006, d = 2.68$). The third hypothesis test had $p = 0.802$, indicating no statistically significant difference between MPC and RL.}

\subsubsection{Temperature Difference Fits}
\label{sec:savings2}

\new{Section \ref{sec:savings1} normalized electricity use by outdoor temperature directly. This approach gives controllers credit for saving energy by reducing indoor temperatures, which might also reduce occupant comfort. This section normalizes electricity use by the outdoor-indoor temperature difference. This approach removes the effect of reducing indoor temperatures, leaving only the effect of operating equipment more efficiently.}

\begin{figure}[ht]
    \centering
    \includegraphics[width=\columnwidth]{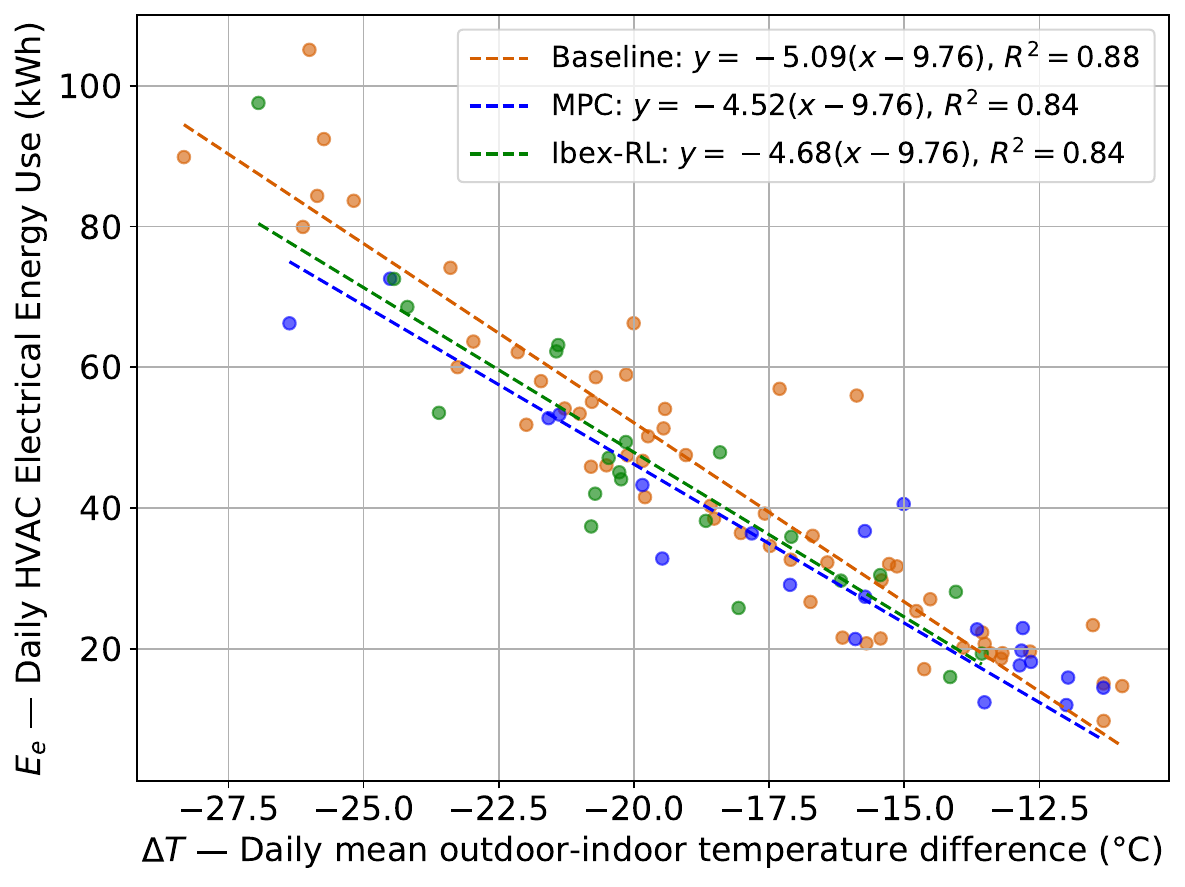}
    \caption{Fits of daily energy use vs. outdoor-indoor temperature difference ($\Delta T$).}
    \label{fig:delta}
\end{figure}

\new{Figure \ref{fig:delta} shows the fits from Eq. \eqref{energyModel} with $x = \overline{T}_{out} - \overline T_{in}$, the daily average outdoor-indoor temperature difference. All control configurations have $R^2 \ge 0.84$. MPC has the least negative slope ($\hat \beta_m = -4.52$ kWh/$^\circ$C), indicating marginally lower daily electricity use than RL ($\hat \beta_r = -4.68$ kWh/$^\circ$C) and significantly lower than the baseline ($\hat \beta_b = -5.09$ kWh/$^\circ$C). Differences in electricity use are larger for larger (absolute) temperature differences.}

\new{The bottom half of Table \ref{tab:controller_performance} summarizes the estimated savings means and 95\% confidence intervals from Eq. \eqref{savingsModel}, alongside the three hypothesis tests from Section \ref{sec:savings1}. The estimated mean relative savings over the baseline for RL and MPC are 8.04\% (95\% CI: 4.14\% to 11.88\%) and 11.28\% (95\% CI: 7.32\% to 15.19\%), respectively. Both effects are statistically significant with $p < 0.001$. The third hypothesis test had $p = 0.251$, indicating no statistically significant difference between MPC and RL.}

\begin{table*}[htb!]
\centering
\caption{\new{Aggregate Energy Savings Summary and Hypothesis Testing.}}
\label{tab:controller_performance}
\begin{tabular}{@{}lcccc@{}}
\toprule
\textbf{Controller} & \textbf{Slope (kWh/$^\circ$C)} & \textbf{Percentage Improvement} & \textbf{$p$-value} & \textbf{Effect Size ($d$)} \\
\midrule
\multicolumn{5}{l}{\textit{Outdoor Temperature Fits (Section \ref{sec:savings1}), $\hat x_0 = 9.78^{\circ}\mathrm{C}$}} \\
\midrule
Baseline & -5.2511 & --- & --- & --- \\
MPC & -4.2901 & 18.13 [4.38 --- 30.90] & 0.006 & 2.68 \\
Ibex-RL & -4.1431 & 20.94 [2.60 --- 38.38] & 0.013 & 2.30 \\
\textit{Diff (MPC - RL)} & --- & \textit{-2.81 [-24.95 --- 19.59]} & \textit{0.802} & \textit{0.25} \\
\midrule
\multicolumn{5}{l}{\textit{Outdoor-indoor Temperature Difference Fits (Section \ref{sec:savings2}), $\hat x_0 = -9.76^{\circ}\mathrm{C}$}} \\
\midrule
Baseline & -5.0898 & --- & --- & --- \\
MPC & -4.5154 & 11.28 [7.32 --- 15.19] & $<0.001$ & 5.61 \\
Ibex-RL & -4.6800 & 8.04 [4.14 --- 11.88] & $<0.001$ & 4.07 \\
\textit{Diff (MPC - RL)} & --- & \textit{3.23 [-2.28 --- 8.76]} & \textit{0.251} & \textit{1.15} \\
\bottomrule
\end{tabular}

\end{table*}

\subsection{Thermal Comfort}

Occupant comfort was monitored via a survey that occupants could complete whenever they felt uncomfortable. Over the 30-day deployment period for RL, occupants submitted this survey on three occasions. The first submission occurred on the second day of deployment, when indoor temperatures dropped and RL quickly recalibrated to prioritize comfort more. The remaining two discomfort reports were submitted on February 13th and February 15th, when respective outdoor temperatures were -11°C and 0°C. In both these later cases, the setpoint and the actual indoor temperature were recorded at 18°C, aligning with the user's stated thermal preference at that time.

To supplement the occupant surveys, we calculated the time-average PPD for all controllers \new{using the same assumptions on unknown parameters such as clothing and activity levels. With those parameters held constant, PPD reduces to a function of the indoor temperature, which was 19.9, 19.1, and 18.3 $^\circ$C on average for the baseline, MPC, and RL controllers, respectively. For comparison, the time-average preference temperature was 19.5 $^\circ$C (18 $^\circ$C for six hours per day and 20 $^\circ$C for the remaining 18 hours).} Table \ref{tab:ppd_stats} shows that the constant-setpoint baseline achieved the best time-average PPD (7.3\%), followed by MPC (9.9\%) and RL (14\%). For reference, practitioners typically consider 10\% PPD acceptable \cite{enescu2017review}.

RL shows higher PPD variability than the baseline or MPC. This is likely due in part to RL's online learning, which leads to changing controller behavior, especially during the initial days of rapid recalibration. A longer RL deployment that gave the model and cost parameters enough time to converge might lead to better comfort performance.

\begin{table}[htbp]
\centering
\caption{PPD Statistics Comparison in \% Values}
\label{tab:ppd_stats}
\resizebox{\linewidth}{!}{%
\begin{tabular}{c lccc}
\toprule 
& Controller: & \textbf{Baseline} & \textbf{MPC} & \textbf{RL} \\
\midrule 
\multirow{3}{*}{\rotatebox[origin=c]{90}{\textbf{\shortstack{Mean \\ (\%)}}}}
& Overall & 7.30  & 9.92  & 13.53 \\
& Day     & 8.03  & 10.59 & 14.06 \\
& Night   & 5.53  & 8.38  & 12.14 \\
\midrule 
\multirow{3}{*}{\rotatebox[origin=c]{90}{\textbf{\shortstack{Std\\Dev \\(\%)}}}}
& Overall & 2.40  & 4.22  & 3.53 \\
& Day     & 2.45  & 4.67  & 3.82 \\
& Night   & 0.88  & 2.30  & 2.07 \\
\midrule 
\multirow{3}{*}{\rotatebox[origin=c]{90}{\textbf{\shortstack{Max \\ (\%)}}}}
& Overall & 21.75 & 24.70 & 29.18 \\
& Day     & 21.75 & 24.70 & 29.18 \\
& Night   & 8.88  & 13.77 & 20.21 \\
\midrule 
\multirow{3}{*}{\rotatebox[origin=c]{90}{\textbf{\shortstack{Min \\ (\%)}}}}
& Overall & 5.00  & 5.08  & 5.40 \\
& Day     & 5.01  & 5.16  & 6.01 \\
& Night   & 5.00  & 5.08  & 5.40 \\

\bottomrule 
\end{tabular}%
} 
\end{table}

\subsection{Deployment Labor}

Deployment labor is considered a primary factor limiting the adoption of advanced HVAC control \cite{henze2025has}, though only a few studies report detailed labor costs \cite{khabbazi2025lessons}. This section provides an empirical account of these costs, measured in engineer-days for the initial, one-time development and commissioning of the specific MPC and RL controllers deployed in this field experiment. To contextualize these labor estimates, each controller was developed by a third-year PhD student, working alongside other projects, as their first real-world deployment of an advanced control system. We therefore use the metric of student-days to transparently represent this effort, which includes the significant learning curve of a skilled but non-expert implementer in a research setting.

The initial MPC deployment was a significant undertaking, requiring approximately 190 student-days. This total can be divided into two main components: roughly 150 student-days for non-recurring, foundational tasks (e.g., cloud database setup, software architecture, sensing infrastructure setup) and the remaining $\sim$40 student-days for house-specific, repeatable tasks. This repeatable portion included fitting the building and equipment models ($\sim$20 days), tuning the control system ($\sim$10 days), and ongoing system maintenance ($\sim$10 days).

The initial development of the RL algorithm took approximately 45 student-days. This work involved an iterative design process to heavily modify Gnu-RL, which is why we were not able to measure task-specific costs. This effort also did not include the non-recurring tasks ($\sim$150 student-days), as the RL deployment used the existing infrastructure from the prior MPC deployment. While the initial tuning for RL was faster, its adaptive nature introduced unique operational challenges. The RL operation was interrupted five times due to external API or measurement system failures, resulting in approximately 7 days of cumulative downtime. Resolving each of these issues required an additional $\sim$0.5 to 2 student-days, highlighting a different type of maintenance burden compared to MPC (which did not require online learning). 

As these were the first MPC and RL deployments for the team, we expect these labor requirements to decrease with experience. However, due to confounding factors like the shared infrastructure and the transfer of knowledge (such as the heat pump COP curve) from the MPC project to the RL project, assessments of future scalability require speculation. To that end, in Section~\ref{sec:scalability} we roughly estimate the recurring deployment costs for another house based on our experience and judgment.

We emphasize that this comparison of deployment labor is not intended as a general verdict on RL versus MPC. The results presented here are specific to the two particular algorithms implemented in this study: Ibex-RL \cite{mulayim2025physics} and the MPC implementation from~\cite{pergantis_field_2024}. The findings are highly contextual, and different outcomes could be expected with alternative RL or MPC implementations. Moreover, the experience and expertise of the engineering team are significant factors that can influence deployment effort and overall performance.

\section{Discussions}
\label{sec:discussions}

This section discusses the key challenges and lessons from our field deployments. A detailed discussion of the lessons learned from MPC implementation can be found in \cite{pergantis_field_2024}.

\subsection{Conflicting Objectives in Imitation Learning}
\label{sec:conflicting}

While differentiable MPC policy underlying Ibex-RL theoretically enables the joint learning of system dynamics and cost parameters, our experiments reveal a challenge in complex real-world systems. This challenge arises from the conflicting requirements of system identification (learning the dynamics model $\theta_{state}$) and imitation learning (learning the quadratic cost parameters $\theta_{cost}$ for a safe start). On one hand, learning an accurate dynamics model for control often requires exciting the system beyond its typical operational range \cite{atam2016control}, which is why MPC used data that included night temperature setbacks. On the other hand, adding excitation would alter the system's behavior, making it impossible to faithfully replicate the existing controller's policy through imitation. In this work, we prioritized imitation and therefore avoided system excitation --- a strategy consistent with our MPC's development, which also relied solely on historical data. Consequently, we knowingly deployed an imperfect system dynamics model, yet Ibex-RL still achieved significant energy savings with minimal comfort compromises.

As a future direction, we propose an alternative method for initializing the quadratic cost parameters offline. This would involve sampling trajectories from historical weather and state data, evaluating them using the reward function $r_t$, and iteratively updating the quadratic cost parameters to maximize cumulative reward. While this offline approach could provide reasonable initial performance aligned with user preferences, it may fail to automatically learn nuanced behaviors like penalizing backup heat usage --- a feature that emerged naturally during imitation learning. Such domain-specific knowledge could be reintroduced via reward shaping, trading slight reductions in scalability for improved initial comfort during deployment.

\subsection{Action Selection Constraints}
\label{sec:choice}

Existing simulation studies often select actions like power output or heat supply without considering real-world controllability constraints. Previous experimental RL implementations in residential settings have employed relatively simple action spaces: \cite{leurs2016beyond} controlled an AC unit through binary on/off commands, \cite{kurte2020evaluating} used high/low setpoints to toggle two AC units, and \cite{svetozarevic2022data} modulated a radiant floor heating valve in a continuous but straightforward manner. Similarly, \cite{montazeri2025fully} used a modular RL agent where the core component decided on a desired temperature change, which was then translated by a separate module into modulating valve openings for radiant heating panels. Our implementation faced more complex constraints. First, we were restricted to setpoint control rather than direct equipment operation, as we lacked access to the heat pump manufacturer's device-level control logic. Second, our action space was further constrained by the physics-based 2R1C model requirements --- all possible actions needed to be physically meaningful and integrable with the system model. These practical limitations resulted in a more challenging but realistic action space compared to previous residential RL experiments.

\subsection{State Representation and Control Mismatch}
\label{ref:temperature}

We selected the return air temperature for the state representation to maintain consistency with the MPC controller, leverage its higher sensor resolution, and better capture whole-house thermal dynamics \cite{Mulayim2024Unmasking}. Both our RL agent and the MPC were constrained to actuating control via thermostat setpoints. This created a fundamental mismatch, as the controllers operated on continuous return air temperature data while the physical thermostat used its own local, quantized sensor with inherent hysteresis. 

Consequently, the thermostat often satisfied local conditions and stopped the HVAC system prematurely. This led to two primary effects for RL: 1) actual energy consumption fell below policy predictions, and 2) online learning was fed with biased state transitions. This outcome was a deliberate engineering trade-off. We accepted the control implementation challenge to gain the high-fidelity data from the return air sensor, which was important for the accurate system identification and imitation learning enable long-term policy improvement.

\subsection{Challenges Working with an Adaptive Controller}
\label{sub:challenges_adaptive}

Deploying an adaptive controller that learns from real-time interactions introduces specific risks regarding data integrity that are not present in standard MPC implementations. In networked systems, communication faults or API drops can prevent a commanded action $u_t$ from being physically implemented. For a static MPC, such drops merely result in a temporary actuation loss without affecting the underlying model. For an online RL agent, however, if the system logs the action as executed, the resulting state transition ($x_t \rightarrow x_{t+1}$) corrupts the training data. Feeding these false transitions into online learning updates introduces inaccurate gradients that can degrade model parameters and lead to unstable policy adaptations. Our deployment highlighted the necessity of implementing strict validation checks --- such as verifying a thermostat setpoint change via an API read-back --- to confirm an action was physically executed before using that transition data for online learning.

\subsection{Safety Assurance in Online Learning} \label{sec:safety} 

Deploying learning-based controllers in occupied housing necessitates strict safety guarantees that standard RL exploration strategies often lack. Our architecture addresses this through multiple layers of safety assurance. First, the underlying differentiable MPC policy structure enforces explicit box constraints on control actions. While our current deployment used thermostat setpoints --- which are inherently bounded by the device's firmware --- this architectural feature is critical for future systems with direct equipment actuation (e.g., modulation signal), where unbounded outputs could cause physical damage. Second, we implemented a deterministic safety layer that strictly prohibited setpoints below 17°C. This hard constraint, which was applied to both the RL and MPC agents, ensured that the house never dropped to unsafe temperatures regardless of the policy's output.

Third, the use of offline imitation learning provided a ``safe but suboptimal'' initialization. Unlike model-free agents that require stochastic exploration to learn --- often leading to erratic behavior in early stages --- our agent began by cloning the historical controller. This ensured that the initial online interactions remained within a reasonable operational envelope while the agent gradually improved its policy. Finally, the proposed formulation inherently supports additional constraints to further protect equipment health. For instance, future work could incorporate penalties on the rate of change of control actions (slew rate) directly into the loss function, effectively damping the policy to prevent high-frequency cycling and extend actuator lifespan.

\subsection{Generalization and Sample Efficiency} \label{sec:generalization}

\new{The field deployment highlights the sample efficiency and behavior under distributional shift.} Learning viable control policies from scratch can require extensive interaction data. Our results demonstrate that the model-based Ibex-RL framework improves this efficiency. Ibex-RL agent adapted \new{its operational strategy to the environment} within 1–2 days of online interaction, \new{enabling rapid deployment without an extended calibration period.}

Regarding distributional shift, model-based RL and MPC approaches can struggle to extrapolate to unseen scenarios \cite{mulayim2025impact, mulayim2025physics}. \new{To mitigate this, our use of a gray-box model (the physics-informed 2R1C network) structurally bounds the state transitions. Even though some identified parameters drifted into unrealistic ranges---such as the negative solar aperture coefficient---the thermodynamic structure of the RC network still prevented unconstrained or erratic control behavior. While the inaccuracies in tracking solar gains inevitably introduced some prediction errors, the structural constraints of the model kept the policy search within a safe operating zone. This allowed the agent to successfully learn the direct impact of its control actions—specifically mastering the minimization of expensive backup heat—and deliver significant energy savings despite the underlying parameter estimation flaws.}

\subsection{Scalability and Recurring Deployment Costs}
\label{sec:scalability}

A key aspect of controller scalability is the recurring labor cost --- the effort required to deploy an already-developed algorithm to a new house. Quantifying this cost is inherently speculative, as it is influenced by numerous project-specific factors like building complexity and the engineer's experience. The following analysis is therefore not a definitive benchmark, but rather an educated estimate based on our experience and judgment, meant to highlight the comparative effort required by our MPC and RL implementations. Furthermore, these estimates reflect a research deployment; a commercial entity would likely invest in automating their software pipeline to significantly reduce these on-boarding costs for a scalable product. This drive for scalability is essential in residential buildings, where high deployment costs can easily outweigh the monetary value of the energy savings.

\new{The following estimates do not account for the initial data collection period and assume the availability of historical data from a pre-existing controller. Under this assumption, we estimate a total of nine engineer-days for the MPC variant implemented here and six engineer-days for the model-based RL variant implemented here, broken down as follows:}

\new{\begin{itemize}
\item \textbf{Shared Infrastructure (four days for both):} Setting up sensors (power and temperature) and establishing an API connection to a smart thermostat requires equivalent effort regardless of the control algorithm.
\item \textbf{MPC Algorithm Configuration (five days):}\footnote{We later deployed the same MPC approach for cooling in the same house \cite{humidity_dchouse}. By reusing the existing model fitting code without tuning, the deployment took only two days; however, the resulting model was not accurate enough to deliver the savings reported here. While using MPC automation toolboxes \cite{de2016toolbox, drgona2023domain} could have reduced the labor gap with RL, it may have resulted in different performance compared to our carefully tuned system.} The MPC variant implemented here used a multi-stage system identification process that requires engineering expertise. The process includes selecting an RC model structure ($\sim$1 day), as well as $\sim$4 days of data pre-processing, fitting parameters, and tuning objective weights.
\item \textbf{RL Algorithm Configuration (two days):} \footnote{This estimate is based on a BOPTEST-gym deployment \cite{arroyo_open-ai_2021}, which involved one day for reconfiguring the model's inputs and outputs and one day for  imitation learning and hyperparameter tuning, implemented in \cite{mulayim2025physics}.}\new{After selecting an RC network structure for Ibex-RL ($\sim$1 day, as in MPC), the second day involves imitation learning, tuning the learning rate $\alpha$ and trade-off parameter $\lambda$ via grid searches, and tuning the discomfort price $w_c$ by simulating temperature tracking using the trained model.}
\end{itemize}}

Ultimately, implementing either MPC or RL for residential HVAC at scale would require bringing per-home deployment costs close to zero. Even one engineer-day of on-boarding effort at typical United States labor and overhead rates could cost more than \$1,000. This is equivalent to several years of energy cost savings from MPC or RL in a typical home. Driving the recurring labor costs close to zero will require a coordinated research effort across several communities. As the control theory community is already moving towards more scalable MPC implementations \cite{drgovna2020all, chinde2022data,drgovna2022differentiable} designed to reduce this labor cost, future experimental studies are needed to test these emerging methods in the field. Transparently reporting on their practical limitations should create a feedback loop for iterative improvement. Concurrently, the software engineering community can address data access hurdles by applying practices like informational requirements and semantics to standardize the data streams and metadata required by these controllers \cite{prakash2024ontologies}. Finally, the RL community must continue to improve the safety and sample efficiency of agents for real-world deployment, with a focus on developing algorithms that do not depend on high-fidelity simulators for training. Together, these parallel efforts can reduce the cost of deploying advanced residential HVAC controllers at scale.

\section{Synthesis, Limitations, and Future Work}
\label{sec:conclusions}

The comparative performance reported here is specific to the particular design choices made for each controller. Both the MPC and RL implementations involved engineering decisions --- such as the MPC's multi-stage parameter fitting process or the assumption of constant thermal mass temperature --- that were guided by physical intuition and field testing. We do not claim that either the MPC or RL implementation is in any sense optimal; rather, this work should be interpreted as a comparative case study of specific, practical implementations of RL and MPC. Yet, the characteristics observed are representative of these particular cases, which in turn reflect broader patterns: MPC's potential for high precision at the cost of engineering effort, and RL's promise of automation coupled with challenges in adaptation. Within this context, our month-long deployment demonstrates that RL is a viable pathway toward scalable HVAC control, but one that involves distinct trade-offs. The RL controller achieved comparable energy savings ($\sim$20.9\% vs. $\sim$18.1\% for MPC) while requiring considerably less engineering overhead for model fitting ($\sim$2 vs. $\sim$5 days). However, this scalability came at the cost of comfort; RL received three occupant discomfort reports and worse PPD scores than MPC. 

The RL and MPC implementations compared in this paper both used a two-level control architecture. RL and MPC each acted as a supervisory controller. The supervisory control action --- the indoor air temperature setpoint --- was sent to the heat pump manufacturer’s device-level controller. This two-level architecture was chosen for scalability: Temperature setpoint adjustments can be sent to most smart thermostats over the Internet via APIs without modifying manufacturer controls or voiding warranties. However, separating the supervisory and device-level controllers is generally suboptimal. Future work could quantify the potential performance improvement from running RL or MPC within the manufacturer’s control system and directly adjusting low-level control variables such as the compressor and fan speeds.

Looking ahead, the deployment also empirically confirmed several other practical challenges for these controllers. Future work could optimize cost parameters offline with a non-quadratic reward function (Section~\ref{sec:conflicting}). Field experiments with adaptive controllers should remove datapoints from online learning batches when chosen actions and implemented actions do not match  (Section~\ref{sub:challenges_adaptive}).
Capacity of the controllers to handle real-world constraints could be improved by better managing discrepancies between the policy's state representation and the physical actuation interface (Sections~\ref{sec:choice} and \ref{ref:temperature}), potentially by modeling the intermediate thermostat control layer. Advancing these techniques to translate complex objectives into tractable learning problems is key to enabling supervisory controllers to deliver both significant energy savings and reliable comfort with minimal engineering effort. This study was also limited to a single testbed. Future work could conduct similar experiments with other equipment configurations, or in other building types or climate zones.

Beyond these specific algorithmic improvements, we point to broader directions for future research. A clear next step for the field is an increased focus on field studies, particularly comparative deployments conducted within the same building, to enable fair evaluation of different control paradigms. Such deployments show the strengths and weaknesses of control strategies, as testing within a single system controls for its unique characteristics while also revealing practical hurdles absent in simulation. Another significant challenge we observed is the limited availability of RL agents suitable for deployment without a pre-existing, high-fidelity simulator. Many existing works focus on model-free agents that require a simulator for training (i.e., model-based acceleration for model-free RL), rather than model-based agents that learn system dynamics automatically, which is a more scalable approach. Indeed, our implementation is, to our knowledge, the first real-world experiment of such a model-based RL algorithm in a residential space, following the only similar deployment (Gnu-RL \cite{chen_gnu-rl_2019}) in the commercial sector. The fact that our own RL controller adopts a structured, model-based approach --- a principle borrowed from MPC to promote safety and interpretability --- may be indicative of a broader trend. This points toward a future defined not by a competition between the MPC and RL paradigms, but by their convergence into hybrid solutions that fuse the automated learning and adaptation of RL within the interpretable, constraint-aware framework of MPC, ultimately enabling more scalable advanced controllers.

\section*{CRediT authorship contribution statement}
Ozan Baris Mulayim: Writing – review \& editing, Writing – original draft, Visualization, Software, Methodology, Investigation, Formal analysis, Data curation, Conceptualization.  Elias N. Pergantis: Writing – review \& editing, Software, Methodology, Data curation. Levi D. Reyes Premer: Writing – review \& editing, Data curation. Bingqing Chen: Writing – review \& editing, Methodology, Conceptualization. Guannan Qu: Writing – review \& editing, Methodology, Conceptualization. Kevin J. Kircher: Writing – review \& editing, Project administration, Methodology, Funding acquisition, Formal analysis, Conceptualization. Mario Bergés: Writing – review \& editing, Project administration, Methodology, Funding acquisition, Formal analysis, Conceptualization.

\section*{Declaration of competing interest}

The authors declare that they have no known competing financial interests or personal relationships that could have appeared to influence the work reported in this paper. Mario Berg\'{e}s holds concurrent appointments as a Professor of Civil and Environmental Engineering at Carnegie Mellon University and as an Amazon Scholar. This paper describes work at Carnegie Mellon University and is not associated with Amazon.

\section*{Data availability}

Data and code will be made available upon publishing.

\section*{Acknowledgments}
The authors would like to thank the occupants of the test-house for their patience and help during testing. The authors would like to gratefully acknowledge the support provided by the Wilton E. Scott Institute for Energy Innovation for Ozan Baris Mulayim. The test-bed creation and maintenance on the Purdue campus was supported through the Center for High-Performance Buildings (project CHPB-26-2024). Elias N. Pergantis and Levi D. Reyes Premer were supported through an ASHRAE (American Society of Heating and Refrigeration Engineers) Grant-in-Aid award. Further, Elias was supported by the Onassis Foundation as one of its scholars and Levi by the National Science Foundation Graduate Research Fellowship (NSF GRF).

\section*{Appendix: Acronyms and Notation}
\label{sec:appendix_nomenclature_short}

This paper used the following acronyms BH: Backup Heat; COP: Coefficient of Performance; HP: Heat Pump; HVAC: Heating, Ventilation, and Air Conditioning; API: Application Programming Interface, KKT: Karush-Kuhn-Tucker; MDP: Markov Decision Process; MPC: Model Predictive Control; PID: Proportional-Integral-Derivative; PPD: Predicted Percentage of Dissatisfied; RC: Resistance-Capacitance; RL: Reinforcement Learning; SVM: Support Vector Machine. Tables \ref{notationTable1} and \ref{notationTable2} summarize the mathematical notation used in this paper.

\begin{table}[htbp]
\centering
\caption{Mathematical Notation (Part 1: Physical System and Parameters)}
\label{tab:notation_physical}
\small
\begin{tabular}{p{0.3\linewidth}p{0.6\linewidth}} 
\toprule
\textbf{Symbol (Units)} & \textbf{Meaning} \\
\midrule

$t, \Delta t$ (h) & Time step index and discretization interval \\

$x, x_t$ ($^\circ$C) & System state vector (e.g., $[T_{in}, T_{mass}]^T$) \\
$u, u_t$ (kW) & Control action vector ($[P_{HP}, P_{BH}]^T$) \\
$d_t$ ($^\circ$C, $^\circ$C, kW/m$^2$) & Disturbance vector ($[T_m, T_{out}, I_{sol}]^T$) \\

$T$ ($^\circ$C) & Temperature: $T_{in}$ (indoor), $T_{out}$ (outdoor), $T_{therm}$ (thermostat), $x_{target}$ (setpoint) \\
$\overline{T}_{out}, \overline{T}$ ($^\circ$C) & Daily mean outdoor/indoor temperature \\
$\Delta T$ ($^\circ$C) & Avg. daily temperature difference ($\overline{T}_{out} - \overline{T}$) \\

$P$ (kW) & Electrical Power: $P_{HP}$ (Heat Pump), $P_{BH}$ (Backup Heater), $P^{max}$ (limits) \\
$\dot{Q}$ (kW) & Heat Rate: $\dot{Q}_c$ (controlled heating), $\dot{Q}_e$ (exogenous gains) \\
$I_{sol}$ (kW/m$^2$) & Solar Irradiance \\
$Q_{day}, E_e$ (kWh) & Daily Energy: $Q_{day}$ (thermal load), $E_e$ (electricity cons.) \\

$C$ (kWh/$^\circ$C) & Building thermal capacitance \\
$R$ ($^\circ$C/kW) & Thermal resistance ($R_m, R_{out}$) \\
$K$ (kW/$^\circ$C) & Global heat loss coefficient ($1/R_{eq}$) \\
$\eta$ (-) & Backup heater efficiency \\
$A_{eff}$ (m$^2$) & Solar aperture coefficient \\
$COP(\cdot)$ (-) & Coefficient of Performance (func. of $T_{out}$) \\
$\theta$ (Mixed) & Parameter Sets: $\theta_{state}$ (physical), $\theta_{cost}$ (weights) \\

\bottomrule
\end{tabular}
\label{notationTable1}
\end{table}

\begin{table}[htbp]
\centering
\caption{Mathematical Notation (Part 2: Dynamics, Control, and Learning)}
\label{tab:notation_control}
\small 
\begin{tabular}{p{0.32\linewidth}p{0.5\linewidth}} 
\toprule
\textbf{Symbol (Units)} & \textbf{Meaning} \\
\midrule

$A, B_u, B_d$ (varies)   & State-space matrices derived from $\theta_{state}$ \\

$O_t, R_t$ & Quadratic cost matrices\\
$O_t, R_{hp}, R_{bh}(-)$ & Quadratic cost parameters ($\theta_{cost}$) \\
$w_d, w_e, w_c$ (\$/kW, \$/kWh, \$/$^\circ$C/h) & Reward weights (demand, energy, discomfort) \\
$J(U_t,x_t)$ (\$) & MPC objective function \\
$r_t$ (\$) & RL instantaneous return \\
$\hat{R}_t (\$)$ & RL estimated undiscounted cumulative reward \\

$\alpha_{\text{imit}}$, $\alpha_{\text{state}}$, $\alpha_{\text{cost}}$ (-)                 & Learning rate hyperparameters (imitation learning, online state updates and online cost parameter updates) \\ 
$\lambda$ (-)            & Regularization coefficient\\
$\gamma$ (-)             & Discount factor (RL) \\
$\mathcal{L}_{\text{imit}}, \mathcal{L}_{state}, \mathcal{L}_{action}$ & Loss functions (imitation, state prediction, action imitation) \\
$G_t$                      & Expected discounted return \\
$Q(x, u)$                  & Action-value function \\
$x_t=\pi_{\theta}(u)$                & Control policy \\
$L$ (steps)                & Lookahead horizon \\

$\beta_0, \beta_1$ (kWh, kWh/$^\circ$C) & Savings function parameters  \\
$\beta_2, \beta_3$ (kWh, kWh/$^\circ$C) & Efficiency function parameters \\

$t, \ell$ (-)            & Discrete time indices \\
$\Delta t$ (h)           & Discrete time step duration \\

\bottomrule
\end{tabular}
\label{notationTable2}
\end{table}

\bibliographystyle{elsarticle-num}
\bibliography{main.bib, sample-base}

@inproceedings{leurs2016beyond,
  title={Beyond theory: Experimental results of a self-learning air conditioning unit},
  author={Leurs, Tim and Claessens, Bert J and Ruelens, Frederik and Weckx, Sam and Deconinck, Geert},
  booktitle={2016 IEEE International Energy Conference (ENERGYCON)},
  pages={1--6},
  year={2016},
  organization={IEEE}
}

@article{svetozarevic2022data,
  title={Data-driven control of room temperature and bidirectional EV charging using deep reinforcement learning: Simulations and experiments},
  author={Svetozarevic, Bratislav and Baumann, Christian and Muntwiler, Simon and Di Natale, Loris and Zeilinger, Melanie N and Heer, Philipp},
  journal={Applied Energy},
  volume={307},
  pages={118127},
  year={2022},
  publisher={Elsevier}
}

@article{liu2006experimental1,
title = {Experimental analysis of simulated reinforcement learning control for active and passive building thermal storage inventory: Part 1. Theoretical foundation},
journal = {Energy and Buildings},
volume = {38},
number = {2},
pages = {142-147},
year = {2006},
issn = {0378-7788},
doi = {https://doi.org/10.1016/j.enbuild.2005.06.002},
author = {Simeng Liu and Gregor P. Henze}
}

@article{khabbazi2025lessons,
title = {Lessons learned from field demonstrations of model predictive control and reinforcement learning for residential and commercial {HVAC}: A review},
journal = {Applied Energy},
volume = {399},
pages = {126459},
year = {2025},
issn = {0306-2619},
doi = {https://doi.org/10.1016/j.apenergy.2025.126459},
author = {Arash J. Khabbazi and Elias N. Pergantis and Levi D. {Reyes Premer} and Panagiotis Papageorgiou and Alex H. Lee and James E. Braun and Gregor P. Henze and Kevin J. Kircher}
}

@inproceedings{mulayim2025physics,
author = {Mulayim, Ozan Baris and Berg\'{e}s, Mario},
title = {Ibex-RL: Interpretable and Scalable Control via Physics-Informed Reinforcement Learning},
year = {2025},
isbn = {9798400719455},
publisher = {Association for Computing Machinery},
address = {New York, NY, USA},
url = {https://doi.org/10.1145/3736425.3770113},
doi = {10.1145/3736425.3770113},
booktitle = {Proceedings of the 12th ACM International Conference on Systems for Energy-Efficient Buildings, Cities, and Transportation},
pages = {192–202},
numpages = {11},
location = {Colorado School of Mines, Golden, CO, USA},
series = {BuildSys '25}
}

@article{zhang2019whole,
  title={Whole building energy model for {HVAC} optimal control: A practical framework based on deep reinforcement learning},
  author={Zhang, Zhiang and Chong, Adrian and Pan, Yuqi and Zhang, Chenlu and Lam, Khee Poh},
  journal={Energy and Buildings},
  volume={199},
  pages={472--490},
  year={2019},
  publisher={Elsevier}
}

@article{liu2006experimental,
  title={Experimental analysis of simulated reinforcement learning control for active and passive building thermal storage inventory: Part 2: Results and analysis},
  author={Liu, Simeng and Henze, Gregor P},
  journal={Energy and buildings},
  volume={38},
  number={2},
  pages={148--161},
  year={2006},
  publisher={Elsevier}
}

@article{WANG2023113026,
title = {Field test of Model Predictive Control in residential buildings for utility cost savings},
journal = {Energy and Buildings},
volume = {288},
pages = {113026},
year = {2023},
issn = {0378-7788},
author = {Dan Wang and Yangzhe Chen and Wei Wang and Cheng Gao and Zhe Wang}
}

@inproceedings{naug2022reinforcement,
  title={Reinforcement learning-based {HVAC} supervisory control of a multi-zone building-{A} real case study},
  author={Naug, Avisek and Quinones--Grueiro, Marcos and Biswas, Gautam},
  booktitle={2022 IEEE conference on control technology and applications (CCTA)},
  pages={1172--1177},
  year={2022},
  organization={IEEE}
}

@inproceedings{hai2025decision,
  title={Decision-Focused Learning for Cost-Efficient and Sustainable Water Treatment Plant Operations},
  author={Hai, Steven Oufan and Ardakanian, Omid},
  booktitle={Proceedings of the 12th ACM International Conference on Systems for Energy-Efficient Buildings, Cities, and Transportation},
  pages={139--149},
  year={2025}
}

@article{wang2024long,
  title={Long-term experimental evaluation and comparison of advanced controls for {HVAC} systems},
  author={Wang, Xuezheng and Dong, Bing},
  journal={Applied Energy},
  volume={371},
  pages={123706},
  year={2024},
  publisher={Elsevier}
}

@inproceedings{chen_gnu-rl_2019,
	address = {New York NY USA},
	title = {Gnu-{RL}: {A} {Precocial} {Reinforcement} {Learning} {Solution} for {Building} {HVAC} {Control} {Using} a {Differentiable} {MPC} {Policy}},
	isbn = {978-1-4503-7005-9},
	shorttitle = {Gnu-{RL}},
	doi = {10.1145/3360322.3360849},
	language = {en},
	urldate = {2023-12-20},
	booktitle = {Proceedings of the 6th {ACM} {International} {Conference} on {Systems} for {Energy}-{Efficient} {Buildings}, {Cities}, and {Transportation}},
	publisher = {ACM},
	author = {Chen, Bingqing and Cai, Zicheng and Bergés, Mario},
	month = nov,
	year = {2019},
	pages = {316--325},
}

@article{de2016toolbox,
  title={Toolbox for development and validation of grey-box building models for forecasting and control},
  author={De Coninck, Roel and Magnusson, Fredrik and {\AA}kesson, Johan and Helsen, Lieve},
  journal={Journal of building performance simulation},
  volume={9},
  number={3},
  pages={288--303},
  year={2016},
  publisher={Taylor \& Francis}
}

@book{sutton1998reinforcement,
  title={Reinforcement learning: An introduction},
  author={Sutton, Richard S and Barto, Andrew G and others},
  volume={1},
  year={1998},
  publisher={MIT press Cambridge}
}

@article{dulac2021challenges,
  title={Challenges of real-world reinforcement learning: definitions, benchmarks and analysis},
  author={Dulac-Arnold, Gabriel and Levine, Nir and Mankowitz, Daniel J and Li, Jerry and Paduraru, Cosmin and Gowal, Sven and Hester, Todd},
  journal={Machine Learning},
  volume={110},
  number={9},
  pages={2419--2468},
  year={2021},
  publisher={Springer}
}

@inproceedings{dong2014real,
  title={A real-time model predictive control for building heating and cooling systems based on the occupancy behavior pattern detection and local weather forecasting},
  author={Dong, Bing and Lam, Khee Poh},
  booktitle={Building Simulation},
  volume={7},
  pages={89--106},
  year={2014},
  organization={Springer}
}

@article{afram2017supervisory,
  title={Supervisory model predictive controller ({MPC}) for residential {HVAC} systems: Implementation and experimentation on archetype sustainable house in Toronto},
  author={Afram, Abdul and Janabi-Sharifi, Farrokh},
  journal={Energy and Buildings},
  volume={154},
  pages={268--282},
  year={2017},
  publisher={Elsevier}
}

@article{bengea2014implementation,
  title={Implementation of model predictive control for an {HVAC} system in a mid-size commercial building},
  author={Bengea, Sorin C and Kelman, Anthony D and Borrelli, Francesco and Taylor, Russell and Narayanan, Satish},
  journal={HVAC\&R Research},
  volume={20},
  number={1},
  pages={121--135},
  year={2014},
  publisher={Taylor \& Francis}
}

@article{sturzenegger2015model,
  title={Model predictive climate control of a swiss office building: Implementation, results, and cost--benefit analysis},
  author={Sturzenegger, David and Gyalistras, Dimitrios and Morari, Manfred and Smith, Roy S},
  journal={IEEE Transactions on Control Systems Technology},
  volume={24},
  number={1},
  pages={1--12},
  year={2015},
  publisher={IEEE}
}

@inproceedings{maasoumy2014model,
  title={Model predictive control approach to online computation of demand-side flexibility of commercial buildings {HVAC} systems for supply following},
  author={Maasoumy, Mehdi and Rosenberg, Catherine and Sangiovanni-Vincentelli, Alberto and Callaway, Duncan S},
  booktitle={2014 American control conference},
  pages={1082--1089},
  year={2014},
  organization={IEEE}
}

@article{wang2023comparison,
  title={Comparison of reinforcement learning and model predictive control for building energy system optimization},
  author={Wang, Dan and Zheng, Wanfu and Wang, Zhe and Wang, Yaran and Pang, Xiufeng and Wang, Wei},
  journal={Applied Thermal Engineering},
  volume={228},
  pages={120430},
  year={2023},
  publisher={Elsevier}
}

@article{arroyo2022comparison,
  title={Comparison of optimal control techniques for building energy management},
  author={Arroyo, Javier and Spiessens, Fred and Helsen, Lieve},
  journal={Frontiers in Built Environment},
  volume={8},
  pages={849754},
  year={2022},
  publisher={Frontiers Media SA}
}

@article{stoffel2023evaluation,
  title={Evaluation of advanced control strategies for building energy systems},
  author={Stoffel, Phillip and Maier, Laura and K{\"u}mpel, Alexander and Schreiber, Thomas and M{\"u}ller, Dirk},
  journal={Energy and Buildings},
  volume={280},
  pages={112709},
  year={2023},
  publisher={Elsevier}
}

@article{drgovna2020all,
  title={All you need to know about model predictive control for buildings},
  author={Drgo{\v{n}}a, J{\'a}n and Arroyo, Javier and Figueroa, Iago Cupeiro and Blum, David and Arendt, Krzysztof and Kim, Donghun and Oll{\'e}, Enric Perarnau and Oravec, Juraj and Wetter, Michael and Vrabie, Draguna L and others},
  journal={Annual Reviews in Control},
  volume={50},
  pages={190--232},
  year={2020},
  publisher={Elsevier}
}

@article{ma2011model,
  title={Model predictive control for the operation of building cooling systems},
  author={Ma, Yudong and Borrelli, Francesco and Hencey, Brandon and Coffey, Brian and Bengea, Sorin and Haves, Philip},
  journal={IEEE Transactions on control systems technology},
  volume={20},
  number={3},
  pages={796--803},
  year={2011},
  publisher={IEEE}
}

@inproceedings{an_clue_2023,
	address = {New York, NY, USA},
	series = {{BuildSys} '23},
	title = {{CLUE}: {Safe} {Model}-{Based} {RL} {HVAC} {Control} {Using} {Epistemic} {Uncertainty} {Estimation}},
	isbn = {9798400702303},
	shorttitle = {{CLUE}},
	doi = {10.1145/3600100.3623742},
	urldate = {2024-09-10},
	booktitle = {Proceedings of the 10th {ACM} {International} {Conference} on {Systems} for {Energy}-{Efficient} {Buildings}, {Cities}, and {Transportation}},
	publisher = {Association for Computing Machinery},
	author = {An, Zhiyu and Ding, Xianzhong and Rathee, Arya and Du, Wan},
	month = nov,
	year = {2023},
	pages = {149--158},
}

@article{pergantis_field_2024,
	title = {Field demonstration of predictive heating control for an all-electric house in a cold climate},
	volume = {360},
	issn = {0306-2619},
	doi = {10.1016/j.apenergy.2024.122820},
	urldate = {2024-10-04},
	journal = {Applied Energy},
	author = {Pergantis, Elias N. and {Priyadarshan} and Theeb, Nadah Al and Dhillon, Parveen and Ore, Jonathan P. and Ziviani, Davide and Groll, Eckhard A. and Kircher, Kevin J.},
	month = apr,
	year = {2024},
	pages = {122820},
}

@inproceedings{tassa2012synthesis,
  title={Control-limited differential dynamic programming},
  author={Tassa, Yuval and Mansard, Nicolas and Todorov, Emo},
  booktitle={2014 IEEE International Conference on Robotics and Automation (ICRA)},
  pages={1168--1175},
  year={2014},
  organization={IEEE}
}

@article{kouvaritakis2016model,
  title={Model predictive control},
  author={Kouvaritakis, Basil and Cannon, Mark},
  journal={Switzerland: Springer International Publishing},
  volume={38},
  number={13-56},
  pages={7},
  year={2016},
  publisher={Springer}
}

@article{jia2019advanced,
  title={Advanced building control via deep reinforcement learning},
  author={Jia, Ruoxi and Jin, Ming and Sun, Kaiyu and Hong, Tianzhen and Spanos, Costas},
  journal={Energy Procedia},
  volume={158},
  pages={6158--6163},
  year={2019},
  publisher={Elsevier}
}

@inproceedings{ding2020mb2c,
author = {Ding, Xianzhong and Du, Wan and Cerpa, Alberto E.},
title = {MB2C: Model-Based Deep Reinforcement Learning for Multi-zone Building Control},
year = {2020},
isbn = {9781450380614},
publisher = {Association for Computing Machinery},
address = {New York, NY, USA},
doi = {10.1145/3408308.3427986},
booktitle = {Proceedings of the 7th ACM International Conference on Systems for Energy-Efficient Buildings, Cities, and Transportation},
pages = {50–59},
numpages = {10},
location = {Virtual Event, Japan},
series = {BuildSys '20}
}

@article{lymperopoulos2020building,
  title={Building temperature regulation in a multi-zone {HVAC} system using distributed adaptive control},
  author={Lymperopoulos, Georgios and Ioannou, Petros},
  journal={Energy and Buildings},
  volume={215},
  pages={109825},
  year={2020},
  publisher={Elsevier}
}

@misc{levine_offline_2020,
	title = {Offline {Reinforcement} {Learning}: {Tutorial}, {Review}, and {Perspectives} on {Open} {Problems}},
	shorttitle = {Offline {Reinforcement} {Learning}},
	url = {http://arxiv.org/abs/2005.01643},
	urldate = {2024-10-21},
	publisher = {arXiv},
	author = {Levine, Sergey and Kumar, Aviral and Tucker, George and Fu, Justin},
	month = nov,
	year = {2020},
	note = {arXiv:2005.01643},
}

@article{amos_differentiable_2019,
  title={Differentiable {MPC} for end-to-end planning and control},
  author={Amos, Brandon and Jimenez, Ivan and Sacks, Jacob and Boots, Byron and Kolter, J Zico},
  journal={Advances in neural information processing systems},
  volume={31},
  year={2018}
}

@article{nakamoto2024cal,
  title={Cal-ql: Calibrated offline {RL} pre-training for efficient online fine-tuning},
  author={Nakamoto, Mitsuhiko and Zhai, Simon and Singh, Anikait and Sobol Mark, Max and Ma, Yi and Finn, Chelsea and Kumar, Aviral and Levine, Sergey},
  journal={Advances in Neural Information Processing Systems},
  volume={36},
  year={2024}
}

@article{kumar2020conservative,
  title={Conservative {Q-learning} for offline reinforcement learning},
  author={Kumar, Aviral and Zhou, Aurick and Tucker, George and Levine, Sergey},
  journal={Advances in Neural Information Processing Systems},
  volume={33},
  pages={1179--1191},
  year={2020}
}

@inproceedings{liu2024adaptive,
  title={Adaptive Policy Regularization for Offline-to-Online Reinforcement Learning in {HVAC} Control},
  author={Liu, Hsin-Yu and Balaji, Bharathan and Gupta, Rajesh and Hong, Dezhi},
  booktitle={Proceedings of the 11th ACM International Conference on Systems for Energy-Efficient Buildings, Cities, and Transportation},
  pages={1--10},
  year={2024}
}

@inproceedings{arroyo_open-ai_2021,
	title = {An {Open}-{AI} gym environment for the {Building} {Optimization} {Testing} ({BOPTEST}) framework},
	doi = {10.26868/25222708.2021.30380},
	author = {Arroyo, Javier and Manna, Carlo and Spiessens, Fred and Helsen, L.},
	month = sep,
	year = {2021},
}

@String{Computing = "Computing" }

@String{Chelsea = "Chelsea" }

@String{Springer = "Springer-Verlag" }

@article{PERGANTIS2025125528,
title = {Protecting residential electrical panels and service through model predictive control: {A} field study},
journal = {Applied Energy},
volume = {386},
pages = {125528},
year = {2025},
issn = {0306-2619},
author = {Elias N. Pergantis and Levi D. {Reyes Premer} and Alex H. Lee and  Priyadarshan and Haotian Liu and Eckhard A. Groll and Davide Ziviani and Kevin J. Kircher},
}

@article{humidity_dchouse,
title = {Humidity-aware model predictive control for residential air conditioning: A field study},
journal = {Building and Environment},
volume = {266},
pages = {112093},
year = {2024},
issn = {0360-1323},
author = {Elias N. Pergantis and Parveen Dhillon and Levi D. Reyes Premer and Alex H. Lee and Davide Ziviani and Kevin J. Kircher},
}

@inproceedings{mulayim2025impact,
author = {Mulayim, Ozan Baris and Berg\'{e}s, Mario},
title = {On the Impact of Simulated Occupancy Behavior Assumptions on Reinforcement Learning for HVAC Controls},
year = {2025},
isbn = {9798400711251},
publisher = {Association for Computing Machinery},
address = {New York, NY, USA},
booktitle = {Proceedings of the 16th ACM International Conference on Future and Sustainable Energy Systems},
pages = {317–321},
numpages = {5},
location = {
},
series = {E-Energy '25}
}

@article{KIM2015279,
title = {Development and experimental demonstration of a plug-and-play multiple {RTU} coordination control algorithm for small/medium commercial buildings},
journal = {Energy and Buildings},
volume = {107},
pages = {279-293},
year = {2015},
issn = {0378-7788},
author = {D. Kim and J.E. Braun and J. Cai and D.L. Fugate}
}

@article{Mulayim2024Unmasking, title={Unmasking the role of remote sensors in comfort, energy, and demand response}, volume={5}, DOI={10.1017/dce.2024.25}, journal={Data-Centric Engineering}, author={Mulayim, Ozan Baris and Severnini, Edson and Bergés, Mario}, year={2024}, pages={e28}}

@article{atam2016control,
  title={Control-oriented thermal modeling of multizone buildings: Methods and issues: Intelligent control of a building system},
  author={Atam, Ercan and Helsen, Lieve},
  journal={IEEE Control systems magazine},
  volume={36},
  number={3},
  pages={86--111},
  year={2016},
  publisher={IEEE}
}

@article{nagy2023ten,
  title={Ten questions concerning reinforcement learning for building energy management},
  author={Nagy, Zoltan and Henze, Gregor and Dey, Sourav and Arroyo, Javier and Helsen, Lieve and Zhang, Xiangyu and Chen, Bingqing and Amasyali, Kadir and Kurte, Kuldeep and Zamzam, Ahmed and others},
  journal={Building and Environment},
  volume={241},
  pages={110435},
  year={2023},
  publisher={Elsevier}
}

@article{kurte2020evaluating,
  title={Evaluating the adaptability of reinforcement learning based {HVAC} control for residential houses},
  author={Kurte, Kuldeep and Munk, Jeffrey and Kotevska, Olivera and Amasyali, Kadir and Smith, Robert and McKee, Evan and Du, Yan and Cui, Borui and Kuruganti, Teja and Zandi, Helia},
  journal={Sustainability},
  volume={12},
  number={18},
  pages={7727},
  year={2020},
  publisher={MDPI}
}

@article{luo2022controlling,
  title={Controlling commercial cooling systems using reinforcement learning},
  author={Luo, Jerry and Paduraru, Cosmin and Voicu, Octavian and Chervonyi, Yuri and Munns, Scott and Li, Jerry and Qian, Crystal and Dutta, Praneet and Davis, Jared Quincy and Wu, Ningjia and others},
  journal={arXiv preprint arXiv:2211.07357},
  year={2022}
}

@article{li2019transforming,
  title={Transforming cooling optimization for green data center via deep reinforcement learning},
  author={Li, Yuanlong and Wen, Yonggang and Tao, Dacheng and Guan, Kyle},
  journal={IEEE transactions on cybernetics},
  volume={50},
  number={5},
  pages={2002--2013},
  year={2019},
  publisher={IEEE}
}

@article{zhang2022safe,
  title={Safe building {HVAC} control via batch reinforcement learning},
  author={Zhang, Chi and Kuppannagari, Sanmukh Rao and Prasanna, Viktor K},
  journal={IEEE Transactions on Sustainable Computing},
  volume={7},
  number={4},
  pages={923--934},
  year={2022},
  publisher={IEEE}
}

@inproceedings{bengea2012model,
  title={Model predictive control for mid-size commercial building {HVAC}: Implementation, results and energy savings},
  author={Bengea, S and Kelman, A and Borrelli, Francesco and Taylor, Russell and Narayanan, Satish},
  booktitle={Second international conference on building energy and environment},
  pages={979--986},
  year={2012}
}

@techreport{drgona2023domain,
  title={Domain Aware Deep-learning Algorithms Integrated with Scientific-computing Technologies (DADAIST)},
  author={Drgona, Jan and Tuor, Aaron R and Koch, James V and Shapiro, Madelyn R and King, Ethan and Vrabie, Draguna L},
  year={2023},
  institution={Pacific Northwest National Laboratory (PNNL), Richland, WA (United States)}
}

@inproceedings{zhang2018deep,
  title={A deep reinforcement learning approach to using whole building energy model for {HVAC} optimal control},
  author={Zhang, Zhiang and Chong, Adrian and Pan, Yuqi and Zhang, Chenlu and Lu, Siliang and Lam, Khee Poh},
  booktitle={2018 Building Performance Analysis Conference and SimBuild},
  volume={3},
  pages={22--23},
  year={2018}
}

@article{henze2025has,
  title={Why has advanced commercial HVAC control not yet achieved its promise?},
  author={Henze, Gregor P and Kircher, Kevin J and Braun, James E},
  journal={Journal of Building Performance Simulation},
  volume={18},
  number={2},
  pages={217--228},
  year={2025},
  publisher={Taylor \& Francis}
}

@inproceedings{prakash2024ontologies,
  title={Ontologies at Work: Analyzing Information Requirements for Model Predictive Control in Buildings},
  author={Prakash, Anand Krishnan and De Andrade Pereira, Flavia and Berg{\'e}s, Mario and Pritoni, Marco and Akinci, Burcu},
  booktitle={Proceedings of the 11th ACM International Conference on Systems for Energy-Efficient Buildings, Cities, and Transportation},
  pages={214--218},
  year={2024}
}

@article{li2014review,
  title={Review of building energy modeling for control and operation},
  author={Li, Xiwang and Wen, Jin},
  journal={Renewable and Sustainable Energy Reviews},
  volume={37},
  pages={517--537},
  year={2014},
  publisher={Elsevier}
}

@article{chinde2022data,
  title={Data-enabled predictive control for building {HVAC} systems},
  author={Chinde, Venkatesh and Lin, Yashen and Ellis, Matthew J},
  journal={Journal of Dynamic Systems, Measurement, and Control},
  volume={144},
  number={8},
  pages={081001},
  year={2022},
  publisher={American Society of Mechanical Engineers}
}

@article{drgovna2022differentiable,
  title={Differentiable predictive control: Deep learning alternative to explicit model predictive control for unknown nonlinear systems},
  author={Drgo{\v{n}}a, J{\'a}n and Ki{\v{s}}, Karol and Tuor, Aaron and Vrabie, Draguna and Klau{\v{c}}o, Martin},
  journal={Journal of Process Control},
  volume={116},
  pages={80--92},
  year={2022},
  publisher={Elsevier}
}

@inproceedings{zhan2023comparing,
  title={Comparing model predictive control and reinforcement learning for the optimal operation of building-{PV}-battery systems},
  author={Zhan, Sicheng and Lei, Yue and Chong, Adrian},
  booktitle={E3S Web of Conferences},
  volume={396},
  pages={04018},
  year={2023},
  organization={EDP Sciences}
}

@article{enescu2017review,
  title={A review of thermal comfort models and indicators for indoor environments},
  author={Enescu, Diana},
  journal={Renewable and Sustainable Energy Reviews},
  volume={79},
  pages={1353--1379},
  year={2017},
  publisher={Elsevier}
}

@article{xu2025efficient,
  title={Efficient and assured reinforcement learning-based building {HVAC} control with heterogeneous expert-guided training},
  author={Xu, Shichao and Fu, Yangyang and Wang, Yixuan and Yang, Zhuoran and Huang, Chao and O’Neill, Zheng and Wang, Zhaoran and Zhu, Qi},
  journal={Scientific reports},
  volume={15},
  number={1},
  pages={7677},
  year={2025},
  publisher={Nature Publishing Group UK London}
}

@article{silvestri2025practical,
  title={Practical deployment of reinforcement learning for building controls using an imitation learning approach},
  author={Silvestri, Alberto and Coraci, Davide and Brandi, Silvio and Capozzoli, Alfonso and Schlueter, Arno},
  journal={Energy and Buildings},
  volume={335},
  pages={115511},
  year={2025},
  publisher={Elsevier}
}

@article{montazeri2025fully,
  title={Fully data-driven and modular building thermal control with physically consistent modeling},
  author={Montazeri, Mina and Remlinger, Carl and Haro, Benjamin Bejar and Heer, Philipp},
  journal={Applied Energy},
  volume={390},
  pages={125770},
  year={2025},
  publisher={Elsevier}
}

@article{lindelof2015field,
  title={Field tests of an adaptive, model-predictive heating controller for residential buildings},
  author={Lindel{\"o}f, David and Afshari, Hossein and Alisafaee, Mohammad and Biswas, Jayant and Caban, Miroslav and Mocellin, Xavier and Viaene, Jean},
  journal={Energy and Buildings},
  volume={99},
  pages={292--302},
  year={2015},
  publisher={Elsevier}
}

@article{blum2019practical,
  title={Practical factors of envelope model setup and their effects on the performance of model predictive control for building heating, ventilating, and air conditioning systems},
  author={Blum, DH and Arendt, K and Rivalin, L and Piette, MA and Wetter, M and Veje, CT},
  journal={Applied Energy},
  volume={236},
  pages={410--425},
  year={2019},
  publisher={Elsevier}
}

\end{document}